\documentstyle[12pt]{article}
\newlength{\bredde}
\def\slash#1{\settowidth{\bredde}{$#1$}\ifmmode\,\raisebox{.15ex}{/}
\hspace*{-\bredde} #1\else$\,\raisebox{.15ex}{/}\hspace*{-\bredde} #1$\fi}
\newcommand{\beq}{\begin{equation}}
\newcommand{\eeq}{\end{equation}}

\def\gtwid{\raise.3ex\hbox{$>$\kern-.75em\lower1ex\hbox{$\sim$}}}
\def\ltwid{\raise.3ex\hbox{$<$\kern-.75em\lower1ex\hbox{$\sim$}}}
\begin{document}
\vspace*{1cm}
\topmargin -0.8cm
\oddsidemargin -0.8cm
\evensidemargin -0.8cm
\headheight 0pt
\headsep 0pt
\topskip 9mm
\title{\Large{
Origin of Antifields in the Batalin-Vilkovisky Lagrangian Formalism}}

\vspace{0.5cm}

\author{
{\sc J. Alfaro}\thanks{Permanent address: Fac. de Fisica, Universidad
Catolica de Chile, Casilla 306, Santiago 22, Chile.} and
{\sc P.H. Damgaard}\\
CERN -- Geneva
}
\maketitle
\vfill
\begin{abstract} The antifields of the Batalin-Vilkovisky Lagrangian
quantization are standard antighosts of certain collective fields.
These collective fields ensure that Schwinger-Dyson equations are
satisfied as a consequence of the gauge symmetry algebra. The
associated antibracket and its canonical structure appear naturally
if one integrates out the corresponding ghost fields. An analogous
Master Equation for the action involving these ghosts follows from
the requirement that the path integral gives rise to the correct
Schwinger-Dyson equations.
\end{abstract}
\vfill
\vspace{4cm}
\begin{flushleft}
CERN--TH-6788/93 \\
January 1993
\end{flushleft}
\vfill
\newpage


\section{Introduction}

When it comes to the quantization of gauge theories in a Lagrangian
formalism, the framework of Batalin-Vilkovisky \cite{Batalin}
appears to be superior to all other available schemes. Not only
is this Batalin-Vilkovisky formalism fairly straightforward to
implement (and, $e.g.$, string field theory can hardly
be done without it \cite{strings}; see also \cite{Verlinde}),
it also gives, for
free, a new canonical structure contained in what is known as
the ``antibracket". With the help of this canonical formalism,
the definition of a gauge-fixed quantum action can be formulated
by means of one equation (and certain subsidiary conditions):
the Master Equation \cite{Batalin}.
In this formalism, it is thus
as if the very definition of a full quantum
field theory can carried through at an algebraic level. We
believe that this is part of its great appeal.

Let us briefly recapitulate the basic ingredients.\footnote{
For some excellent recent reviews, see ref. \cite{review}.} Start
with a set of fields $\phi^A(x)$ of given Grassmann parity
(statistics) $\epsilon(\phi^A)=\epsilon_A$, and then introduce for
each field a corresponding antifield $\phi^*_A$ of opposite Grassmann
parity $\epsilon(\phi^*_A)=\epsilon_A+1$.
The fields and antifields are taken to be canonically
conjugate,
\beq
(\phi^A,\phi^*_B) = \delta^A_B~,~~~ (\phi^A,\phi^B) =
(\phi^*_A,\phi^*_B) ~=~ 0 ,
\eeq
within a certain graded bracket structure $(\cdot,\cdot)$,
the antibracket:
\beq
(F,G) = \frac{\delta^r F}{\delta\phi^A}\frac{\delta^l G}
{\delta\phi^*_A}
-\frac{\delta^r F}{\delta\phi^*_A}\frac{\delta^l G}{\delta\phi^A}~.
\eeq
The subscripts $l$ and $r$ denote left and right differentiation,
respectively. The summation over indices $A$ includes an integration
over continuous variables such as space-time $x$, when required.

This antibracket is statistics-changing in the sense that
\beq
\epsilon[(F,G)] = \epsilon(F) + \epsilon(G) + 1 ,
\eeq
and satisfies the following exchange relation:
\beq
(F,G) ~=~ -(-1)^{(\epsilon(F)+1)(\epsilon(G)+1)}(G,F) ~.
\eeq
Furthermore, one may verify that the antibracket acts as a derivation
of the kind
\begin{eqnarray}
(F,GH) &=& (F,G)H + (-1)^{\epsilon(G)(\epsilon(F)+1)}G(F,H) \cr
(FG,H) &=& F(G,H) + (-1)^{\epsilon(G)(\epsilon(H)+1)}(F,H)G  ~,
\end{eqnarray}
and satisfies a Jacobi identity of the form
\beq
(-1)^{(\epsilon(F)+1)(\epsilon(H)+1)}(F,(G,H)) +
{\mbox{\rm cyclic perm.}} = 0 ~.
\eeq
Some simple consequences of these relations are that $(F,F) = 0$ for
any Grassmann odd $F$, and $(F,(F,F)) = ((F,F),F) = 0$ for any $F$.

The antifields $\phi^*_A$ are also given definite ghost numbers
$gh(\phi^*_A)$, related to those of the fields $\phi^A$:
\beq
gh(\phi^*_A) = - gh(\phi^A) - 1 ~.
\eeq
The absolute value of the ghost number can be fixed by requiring that
the action carries ghost number zero.

The Batalin-Vilkovisky quantization prescription can now be formulated
as follows. First solve the equation
\beq
\frac{1}{2}(W,W) ~=~ i\hbar \Delta W
\eeq
where
\beq
\Delta ~=~ (-1)^{\epsilon_A+1}\frac{\delta^r}{\delta\phi^A}
\frac{\delta^r}{\delta\phi^*_A} ~.
\eeq
This $W$ will be the ``quantum action", presumed expandable in powers of
$\hbar$:
\beq
W = S + \sum_{n=0}^{\infty} \hbar^n M_n ,
\eeq
and a boundary condition is that $S$ in eq. (10) should coincide with
the classical action when all antifields are removed, $i.e.$, after
setting $\phi^*_A = 0$. One can solve for the additional $M_n$-terms
through a recursive procedure, order-by-order in an $\hbar$-expansion.
To lowest order in $\hbar$ this is the Master Equation:
\beq
(S,S) ~=~ 0 ~.
\eeq
Similarly, one can view eq. (8) as the full ``quantum Master Equation".

A correct path integral prescription for the
quantization of the classical theory $S[\phi^A,\phi^*_A=0]$ is that
one should find an appropriate ``gauge fermion" $\Psi$ (the precise
properties of which need not concern us yet) such that the partition
function is given by
\beq
{\cal{Z}} ~=~ \int [d\phi^A][d\phi^*_A]\delta(\phi^*_A -
\frac{\delta^r \Psi}{\delta \phi^A})\exp\left[\frac{i}{\hbar}W\right]~.
\eeq
This prescription guarantees gauge-independence of the $S$-matrix of
the theory. The ``extended action" $S \equiv S_{ext}[\phi^A,\phi^*_A]$
is a solution of the Master Equation (11),
and can been given by an expansion in powers of antifields
\cite{Batalin}. After the elimination of the antifields by the
$\delta$-function constraint in eq. (12), one can verify that the
action is invariant under the usual BRST symmetry, which we will
here denote by $\delta$.

While this formalism thus in a beautiful way encompasses the usual
Lagrangian BRST quantization (as formulated in, $e.g.$, ref.
\cite{Kugo}), it in a still mysterious way also seems to supersede it.
Is there any rationale behind this?

But there are other, simpler, questions as well:
\begin{itemize}
\vspace{0.2cm}
\item Why do we suddenly have to effectively
double the number of fields involved (by introducing the antifields), and
then immediately remove these fields again through the introduction of
a gauge fermion $\Psi$, and the $\delta$-function prescription of eq.(12)?
\vspace{0.2cm}
\item Is such a $\delta$-function prescription for removing the
antifields the most general?
\vspace{0.2cm}
\item In this formalism, the full quantum action will in general
contain all powers of $\hbar$. On the surface, this seems different
from more conventional BRST quantization methods. What is the
origin of these quantum corrections?
\vspace{0.2cm}
\item Denoting the usual BRST transformation by $\delta$,
the Batalin-Vilkovisky formalism induces a cohomology based
not on $\delta$ but on the ``quantum deformation" acting on both
fields and antifields, which is defined by the nilpotent
$\sigma = \delta - i\hbar\Delta$.
Such a split into classical and quantum
mechanical parts of symmetries is unnatural; from where does it arise?
\vspace{0.2cm}
\item Is it possible to give a step-by-step Lagrangian derivation
of the quantization principle?
\end{itemize}
\vspace{0.2cm}

We shall try to answer these questions one by one.\footnote{For a
geometrical interpretation of some of these issues, see ref.
\cite{Witten}.} In the process of doing
so, we will uncover and derive in a simple manner the Batalin-Vilkovisky
formalism starting from two basic ingredients: A) Standard BRST Lagrangian
quantization, and B) The requirement that the most general Schwinger-Dyson
equations of the full quantum theory follow from a symmetry principle.
We shall throughout, unless otherwise stated, assume that when
ultraviolet regularization is required, a suitable regulator
which preserves the relevant BRST symmetry exists.

Schwinger-Dyson equations will thus play a crucial r\^{o}le in this
analysis. The idea is to enlarge the usual BRST
symmetry in precisely such a way that both the usual gauge-symmetry Ward
Identities {\em and} the most general Schwinger-Dyson equations both
follow from the same BRST Ward Identities. The way to do this is known
\cite{us};
it is a special case of collective field transformations that can
be used to gauge arbitrary symmetries \cite{us1}. In fact, almost all
the needed ingredients can be found in ref. \cite{us1}, but we have
made an effort towards writing this paper in a self-contained manner.
No familiarity with the Batalin-Vilkovisky formalism beyond what has
been sketched above is required.

We start in section 2 with the simplest case possible, that of a quantum
field theory without any internal gauge degrees of freedom. This will
nevertheless allow us to derive most of the algebraic structure needed
subsequently: the antibracket, the gauge-fixing principle, the origin
of a ``quantum" BRST algebra, and the Master Equation itself. One of
the conclusions we shall draw is that the required quantum corrections
to the BRST generator can be seen as an
artifact of having integrated out only {\em one} ghost field instead
of the usual procedure of either integrating out both ghosts and
antighosts simultaneously, or leaving them both. In section
3 we turn to a more non-trivial example: A gauge theory with Yang-Mills
symmetry. We show here how the same principle of requiring
Schwinger-Dyson equations to follow from the BRST algebra leads to
the additional ingredients of the Batalin-Vilkovisky formalism: The
extended action, the general transformation laws for the antifields, and
the gauge fixing as a consequence of a canonical transformation (within
the antibracket) of both fields and antifields. Similarly, the
so-called non-minimal action can be derived in a completely analogous
manner. This section 3 is highly technical, and some readers may wish
to jump immediately to section 4 where the scheme is explained in a
more condensed manner, and where also various generalizations are
discussed. In particular,
cases that can not be straightforwardly derived by means of collective
fields (as, $e.g.$, theories with open gauge algebras), require special
care. Demanding that Schwinger-Dyson equations follow as BRST Ward
Identities for such theories will in general correspond to actions and
functional measures that are not separately invariant under the BRST
transformations. As a consequence, the analogous Master Equation will
contain a quantum correction, -- the origin of the quantum Master
Equation of Batalin and Vilkovisky. In this precise sense, the quantum
Master Equation is a necessary and sufficient requirement for a
consistent quantization of the theory, even before any gauge fixing.
Section 5 contains our conclusions.
We have listed our conventions and some useful formulae
in an appendix.

\section{No gauge Symmetries}

Consider a quantum field theory based on an action $S[\phi^A]$ without
any internal gauge symmetries. Such a quantum theory can be described
by a path integral\footnote{Since one of the curious properties of
the Batalin-Vilkovisky formalism is the combination of
classical and quantum parts in the BRST cohomology, it is useful
to keep track of factors of $\hbar$.}
\beq
{\cal{Z}} ~=~ \int [d\phi^A] \exp\left[\frac{i}{\hbar} S[\phi^A]\right] ,
\eeq
and the associated generating functional. There is no need to introduce
sources for the present discussion.

Equivalently, such a quantum field theory is believed to be entirely
described by the solution of the corresponding Schwinger-Dyson
equations, once appropriate boundary conditions have been imposed.
Schwinger-Dyson equations are therefore quantum mechanically {\em exact}
statements about the theory. At the path integral level they follow
from invariances of the measure. Let us for simplicity consider the
case of a flat measure which is invariant under arbitrary local shifts,
$\phi^A(x) \to \phi^A(x) + \varepsilon^A(x)$. We can gauge this symmetry
by means of collective fields $\varphi^A(x)$: Suppose we transform the
original field as
\beq
\phi^A(x) \to \phi^A(x) - \varphi^A(x) ,
\eeq
then the transformed action $S[\phi^A-\varphi^A]$ is trivially invariant
under the local gauge symmetry
\beq
\delta \phi^A(x) = \Theta(x)~,~~~~ \delta \varphi^A(x) = \Theta(x) ,
\eeq
and the measure for $\phi^A$ in eq. (13) is also invariant. We next
integrate over the collective field in the transformed path integral,
using the same flat measure.
The integration is of course very formal since it will include the whole
volume of the gauge group.\footnote{This situation is no different from
usual path integral manipulations of
gauge theories.} To cure this problem, we gauge-fix in the standard
BRST Lagrangian manner \cite{Kugo}. That is, we add to the transformed
Lagrangian a BRST-exact term in such a way that the local gauge symmetry
is broken. In this case an obvious BRST multiplet consists of a
ghost-antighost pair $c^A(x), \phi^*_A(x)$, and a Nakanishi-Lautrup
field $B_A(x)$:
\begin{eqnarray}
\delta \phi^A(x) &=& c^A(x) \cr
\delta \varphi^A(x) &=& c^A(x) \cr
\delta c^A(x) &=& 0 \cr
\delta \phi^*_A(x) &=& B_A(x) \cr
\delta B_A(x) &=& 0 .
\end{eqnarray}

No assumptions will be made as to whether $\phi^A$ are of odd or even
Grassmann parity. We assign the usual ghost numbers to the new fields,
\beq
gh(c^A) = 1~,~~~ gh(\phi^*_A) = -1~,~~~ gh(B_A) = 0,
\eeq
and the operation $\delta$ is statistics-changing. The rules for
operating with $\delta$ are given in the Appendix.

Let us choose to gauge-fix the transformed action
by adding to the Lagrangian a term of the form
\beq
-\delta[\phi^*_A(x)\varphi^A(x)] = (-1)^{\epsilon(A)+1}B_A(x)
\varphi^A(x) - \phi^*_A(x)c^A(x) .
\eeq
The partition function is now again well-defined:
\beq
{\cal{Z}} ~=~ \int [d\phi][d\varphi][d\phi^*][dc][dB]
\exp\left[\frac{i}{\hbar}\left(S[\phi-\varphi] - \int dx\{
(-1)^{\epsilon(A)}B_A(x)\varphi^A(x) + \phi^*_A(x)c^A(x)\}
\right)\right] .
\eeq
Since the collective field has just been gauge fixed to zero, it
may appear useful to integrate both it and the field $B_A(x)$ out.
We are then left with
\begin{eqnarray}
{\cal{Z}} &=& \int [d\phi^A][d\phi^*_A][dc^A] \exp\left[\frac{i}{\hbar}
S_{ext}\right] \cr
S_{ext} &=& S[\phi^A] - \int dx \phi^*_A(x)c^A(x) ~,
\end{eqnarray}
which obviously coincides with the original expression (13) apart from
the trivially decoupled ghosts. But the remnant BRST symmetry is
still non-trivial: We find it in the usual way by substituting
for $B_A(x)$ its equation of motion. This gives
\begin{eqnarray}
\delta \phi^A(x) &~=~& c^A(x) \cr
\delta c^A(x) &~=~& 0 \cr
\delta \phi^*_A(x) &~=~& - \frac{\delta^l S}{\delta \phi^A(x)} .
\end{eqnarray}

The functional measure is also invariant under this symmetry,
according to our assumption about the measure for $\phi^A$, and
assuming a flat measure for $\phi^*_A$ as well.
The Ward Identities following from this symmetry are the
seeked-for Schwinger-Dyson equations. Consider, for example, the
identity $0 = \langle \delta\{\phi^*_A(x)F[\phi^A]\}\rangle$, where
we have chosen $F$ to depend only on $\phi^A$ just to ensure that
the whole object carries overall ghost number zero (more general $F$'s
can of course also be considered). After integrating over both
ghosts $c^A$ and antighosts $\phi^*_A$, this Ward Identity can be
written
\beq
\langle \frac{\delta^lF}{\delta\phi^A(x)} +
\left(\frac{i}{\hbar}\right)\frac{\delta^lS}{\delta\phi^A(x)}
F[\phi^A] \rangle = 0 ~,
\eeq
that is, precisely the most general Schwinger-Dyson equations for
this theory. The symmetry (21) can be viewed as the BRST Schwinger-Dyson
algebra.

Consider now the equation that expresses BRST invariance of the extended
action $S_{ext}$:
\begin{eqnarray}
0 = \delta S_{ext} &=& \int dx \frac{\delta^r S_{ext}}{\delta\phi^A(x)}
c^A(x) - \int dx \frac{\delta^r S_{ext}}{\delta\phi^*_A(x)}
\frac{\delta^l S}{\delta\phi^A(x)} \cr
&=& \int dx \frac{\delta^r S_{ext}}{\delta\phi^A(x)} c^A(x)
- \int dx \frac{\delta^r S_{ext}}{\delta\phi^*_A(x)}
\frac{\delta^l S_{ext}}{\delta\phi^A(x)}~.
\end{eqnarray}
In the last line we have used the fact that $S$ differs from
$S_{ext}$ by a term independent of $\phi^A$. Using the notation of
the antibracket (2), this is seen to correspond to a Master Equation
of the form
\beq
\frac{1}{2}(S_{ext},S_{ext}) = - \int dx \frac{\delta^r S_{ext}}
{\delta\phi^A(x)}c^A(x) ~.
\eeq

The ghosts $c^A$ play the r\^{o}le of spectator fields in the
antibracket. But their appearance on the r.h.s. of the Master Equation
ensures that the solution $S_{ext}$ will contain these fields.

If one prefers to view the Master Equation (24) as more fundamental,
one can take this equation as the starting point. To find a perhaps
more general solution, let us assume that it can be written as an
expansion in ghosts and antighosts. Since $S_{ext}$ should have ghost
number zero, we try a general expansion of the form
\beq
S_{ext}[\phi^A,\phi^*_A,c^A] = S[\phi^A] + \sum_{n=1}^{\infty}
a_n \phi^*_{A_{1}}\ldots\phi^*_{A_{n}}c^{A_{1}}\ldots c^{A_{n}}
\eeq
with unknown coefficients $a_n$. We have imposed the boundary condition
that $S_{ext}[\phi^A,0,0] = S[\phi^A]$. The most general expansion in
ghosts and antighosts involves $\phi^A$-dependent coefficients
$a_n$, but then the relevant Schwinger-Dyson BRST algebra is not
guaranteed to leave the functional measure invariant. The Master
Equation (24) will then have to be changed. There is no need to enter
into a discussion of these complications at this stage (they will
resurface when we discuss the general procedure in section 4). We shall
therefore for the moment content ourselves with the expansion (25).
It appears in any case at this stage superficially required to have
$\phi^A$-independent
coefficients $a_n$ if we insist on obtaining Schwinger-Dyson equations
for the original theory described by $S[\phi^A]$, since otherwise
the manipulation in eq. (23) would not be valid.
Plugging (25) into the Master Equation (24), we can now
compare order by order in the number of ghost and antighost fields.
This immediately leads to $a_1 = 1$, which in turn implies that all
higher coefficients vanish, $i.e.$,
\beq
a_1 = 1~, ~~~~~a_n = 0 ~~~~{\mbox{\rm for all}}~
n > 1~.
\eeq

In this case of no internal gauge symmetries,
the action (20) is therefore the unique solution to the Master
Equation (24) with the boundary condition $S_{ext}[\phi^A,
\phi^*_A=0,c^A=0] = S[\phi^A]$ and with the new ghosts and
antighosts decoupled from the classical action.

The extended action of Batalin and Vilkovisky does however
not coincide with $S_{ext}$ as defined above. For one thing,
the action (20) contains the new ghost fields $c^A$ that are not
present in the Batalin-Vilkovisky formalism.
But suppose we integrate {\em only} over these ghosts $c^A(x)$,
without integrating over the corresponding antighosts $\phi^*_A(x)$.
Then the partition function reads
\beq
{\cal{Z}} = \int [d\phi^A][d\phi^*_A]\delta\left(\phi^*_A\right)
\exp\left[\frac{i}{\hbar}S[\phi^A]\right] .
\eeq

What has happened to the BRST algebra? As far as symmetry
transformations are concerned, the usual rule for replacing
non-propagating fields which are being integrated out, is to use the
corresponding equations of motion. This recipe is in general
only correct for bosonic Gaussian integrals.\footnote{The
reader may well ask why we were allowed to use a similar
simple rule in the previous example, eq. (19), where we
integrated out a field ($B_A$) that only appeared linearly in the
action. The reason is that we simultaneously integrated out
the collective field $\varphi^A$. Had we chosen to keep
$\varphi^A$, the substitution rule would indeed be more
complicated, and in fact entirely analogous to the recipe we
shall provide above.} For the present case of a ghost field $c^A$
appearing linearly in the action before being integrated out, we
can derive the correct substitution rule as follows.
First, we should really phrase the question in a more precise
manner. What we need to know is how to replace $c$ inside
the path integral, $i.e.$, inside Green functions.
(In the action it is of course not present, having been integrated
out). This will automatically give us the correct transformation
rules for those fields that are not integrated out. Consider
the identity
\beq
\int [dc] F(c^B(y))\exp\left[-\frac{i}{\hbar}\int dx
\phi^*_A(x)c^A(x)
\right] = F\left(i\hbar\frac{\delta^l}{\delta \phi^*_B(y)}\right)
\exp\left[-\frac{i}{\hbar}\int dx\phi^*_A(x)c^A(x)\right] ~.
\eeq

Eq. (28) teaches us that it is not enough to replace $c$ by its
equation of motion ($c(x)=0$); a ``quantum correction" in the
form of the operator $\hbar\delta/\delta\phi^*$ must be added as
well. The appearance of this operator is the final step
towards unravelling the canonical structure in the formalism of
Batalin and Vilkovisky. It also shows that even in this trivial
case we have to include ``quantum corrections" to BRST symmetries
if we insist on integrating out only one ghost field, while
keeping its antighost.

It is important that the operator $\hbar\delta^l/\delta\phi^*$
in eq. (28) always acts on the integral (really a $\delta$-function)
to its right.

We now make this replacement, having always in mind that it is only
meaningful inside the path integral. For the BRST transformation
itself we get, upon one partial integration
\begin{eqnarray}
\delta\phi^A(x) &=& i\hbar(-1)^{\epsilon_A}
\frac{\delta^r}{\delta\phi^*_A(x)} \cr
\delta\phi^*_A(x) &=& - \frac{\delta^l S}{\delta\phi^A(x)} ~.
\end{eqnarray}
We know from our derivation that this transformation leaves at least
the combination of measure and action invariant. As a check, if we
consider the same Ward Identity as above, based on $0 = \langle
\delta\{\phi^*_A(x)F[\phi^A]\}\rangle$, we recover the
Schwinger-Dyson equation (22).

The original $S[\phi^A]$ can in this case be identified with the
extended action of Batalin and Vilkovisky, and the antighost
$\phi^*_A$ is the antifield corresponding to $\phi^A$. Because
there are no internal gauge symmetries, the extended action turns
out to be independent of the antifields. Although our $S_{ext}$
of eq. (20) cannot
be identified with the extended action of Batalin-Vilkovisky, the
Master Equation derived in eq. (23) is of course very similar
to their corresponding Master Equation.
Writing eq.(23) in terms of the antibracket is, as follows from the
derivation,
a little forced. It is done in order to facilitate the
comparison.

Finally, it only remains to be seen what has happened to this
bracket structure after having integrated out the ghost. It is
convenient to keep the same notation as before, so that in
this case the ``extended" action is trivially equal to the
original action: $S_{ext} = S[\phi^A]$. Let us then again
consider the variation of an arbitrary functional $G$, this time
only a function of $\phi^A$ and $\phi^*_A$. Inside the path
integral (and only there!) we can represent the variation of $G$
as:
\beq
\delta G[\phi^A,\phi^*_A] = \int dx\frac{\delta^r G}{\delta\phi^A(x)}
\left[\frac{\delta^l S_{ext}}{\delta\phi^*_A(x)} +
(i\hbar)(-1)^{\epsilon_A}\frac{\delta^r}
{\delta\phi^*_A(x)}\right] - \int dx\frac{\delta^r G}
{\delta\phi^*_A(x)}\frac{\delta^l S_{ext}}{\delta\phi^A(x)}~,
\eeq
where the derivative operator no longer acts on the
$\delta$-function of $\phi^*_A$. We have kept the term proportional
to $\delta^l S_{ext}/\delta\phi^*_A$, even though $S_{ext}$ in this
simple case is independent of $\phi^*_A$ (and that term therefore
vanishes). It comes from the partial integration with respect to
the operator $\delta^l/\delta^*_A$, and is there in general when
$S_{ext}$ depends on $\phi^*_A$ (see section 4.2).

This equation precisely describes the ``quantum deformation" of the
classical BRST charge, as it occurs in the Batalin-Vilkovisky
framework:
\beq
\delta G = (G,S_{ext}) - i\hbar \Delta G ~.
\eeq
The BRST operator in this form is often denoted by $\sigma$.

We have here introduced
\beq
\Delta \equiv (-1)^{\epsilon_A+1}\frac{\delta^r\delta^r}
{\delta\phi^*_A(x)
\delta\phi^A(x)}
\eeq
which is identical to the operator (9) of Batalin and
Vilkovisky. Again, this term arises as a consequence of the
partial integration which allows us to expose the operator
$\delta/\delta\phi^*_A(x)$ that otherwise (in a correct, but
inconvenient representation) acts only on the
$\delta$-functional $\delta(\phi^*_A)$. In other words, the
$\delta$-function constraint on $\phi^*_A(x)$ is now considered as part
of the functional measure for $\phi^*_A$.

Precisely which properties of the partially-integrated extended action
are then responsible for the canonical structure behind the
Batalin-Vilkovisky formalism? As we have seen, the crucial ingredients
come from integrating out the Nakanishi-Lautrup fields $B_A$ and the
ghosts $c^A$. Integrating out $B_A$ changes the BRST variation of the
antighosts $\phi^*_A$ into $-\delta^l S_{ext}/\delta\phi^A$.
{\em This is
an inevitable consequence of introducing collective fields as shifts
of the original fields (and hence enforcing Schwinger-Dyson equations),
and then gauge-fixing them to zero}. But this only provides half of the
canonical structure, making, loosely speaking, $\phi^A$ canonically
conjugate to $\phi^*_A$, but not vice versa. The rest is provided by
integrating over the ghosts $c^A$; at the linear level it changes the
BRST variation of the fields $\phi^A$ themselves
into $\delta^l S_{ext}/\delta\phi^*_A$. {\em This in turn is again an
inevitable consequence of having introduced the collective fields as
shifts, and then having gauge-fixed them to zero.}\footnote{Gauge-fixing
the collective fields to zero implies linear couplings to
the auxiliary fields and ghosts, respectively. This is one of the central
properties of the extended action that leads to the canonical structure,
and to the fact that the extended action $S_{ext}$ itself is the
(classical) BRST generator. However, this does not exclude the
possibility that different gauge fixings of the shift symmetries could
produce a more general formalism.}

The latter operation is,
however, complicated by the fact that the fields $\phi^*_A$ which are
fixed in the process of integrating out $c^A$ are chosen to remain
in the path integral. This makes it impossible to discard the ``quantum
correction" to $\delta S_{ext}/\delta\phi^*_A$. So, in fact, if one
insists on keeping these antighosts $\phi^*_A$, now seen as canonically
conjugate partners of $\phi^A$, the simple canonical structure is in this
sense never truly realized.\footnote{It will not help to
integrate further, and remove the antifields as well. Then the canonical
structure is lost yet again, and the ``Master Equation" is then simply
the condition that the gauge-fixed action be invariant under the
BRST symmetry of the internal gauge invariances.}

We have seen these features only in what is the trivial
case of no internal gauge symmetries. But as we shall show in the
following two sections, they hold in greater generality.

\section{Gauge Theories: Yang-Mills}

Before giving a more general presentation of the equivalence between
the Batalin-Vilkovisky scheme and the collective field approach,
let us first describe another simple example: Yang-Mills theory. It will
turn out to contain most of the still missing ingredients.

We start with the pure Yang-Mills action $S[A_\mu]$, and define a
covariant derivative with respect to $A_\mu$ as
\beq
D^{(A)}_\mu ~\equiv~ \partial_\mu - [A_\mu,~] .
\eeq
Next, we introduce the collective field $a_\mu$ which will enforce
Schwinger-Dyson equations as a result of the BRST algebra:
\beq
A_\mu(x) \to A_\mu(x) - a_\mu(x) .
\eeq
In comparison with the previous example, the only new aspect
here is that the transformed action $S[A_\mu-a_\mu]$ actually has two
independent gauge symmetries. Because of the redundancy introduced by
the collective field, we can write the two symmetries in different ways.
To make contact with the Batalin-Vilkovisky formalism, we will choose
a very particular version, the one corresponding to
\begin{eqnarray}
\delta A_\mu(x) &=& \Theta_\mu(x) \cr
\delta a_\mu(x) &=& \Theta_\mu(x) - D^{(A-a)}_\mu\varepsilon(x) ,
\end{eqnarray}
which also shows the need for being careful in defining what we mean
by a covariant derivative. (The general principle is that we choose
the original gauge symmetry of the original field to be carried
entirely by the collective field; the transformation of the original
gauge field is then always just a shift.)
Although $\Theta(x)$ includes arbitrary
deformations, it only leaves the transformed $\em field$ invariant,
while of course the action is also invariant under Yang-Mills gauge
transformations of this transformed field itself. Hence the need for
including two independent gauge transformations.

We now gauge-fix these two gauge symmetries, one at a time, in the
standard BRST manner. As a start, we introduce a suitable
multiplet of ghosts and auxiliary fields. We need one Lorentz
vector ghost $\psi_\mu(x)$ for the shift symmetry of $A_\mu$,
and one Yang-Mills ghost $c(x)$. These are of course Grassmann
odd, and both carry the same ghost number
\beq
gh(\psi_\mu) ~=~ gh(c) ~=~ 1 .
\eeq

Next, we gauge-fix the shift symmetry of $A_\mu$ by removing the
collective field $a_\mu$. This leads us to introduce a corresponding
antighost $A^*_\mu(x)$, Grassmann odd, and an auxiliary field
$b_\mu(x)$, Grassmann even. They have the usual ghost number
assignments,
\beq
gh(A^*_\mu) ~=~ -1~,~~~~ gh(b_\mu) ~=~ 0 ~,
\eeq
and we now have the nilpotent BRST algebra
\begin{eqnarray}
\delta A_\mu(x) &=& \psi_\mu(x) \cr
\delta a_\mu(x) &=& \psi_\mu(x) - D^{(A-a)}_\mu c(x) \cr
\delta c(x) &=& -\frac{1}{2}[c(x),c(x)] \cr
\delta \psi_\mu(x) &=& 0 \cr
\delta A^*_\mu(x) &=& b_\mu(x) \cr
\delta b_\mu(x) &=& 0 .
\end{eqnarray}

Fixing $a_\mu(x)$ to zero is achieved by adding a term
\beq
-\delta[A^*_\mu(x)a^\mu(x)] = - b_\mu(x)a^\mu(x) -
A^*_\mu(x)\{\psi^\mu - D_{(A-a)}^\mu c(x)\}
\eeq
to the Lagrangian. We shall follow the usual rule of only starting
to integrate over
{\em pairs} of ghost-antighosts in the partition function. With this
rule we shall still keep $c(x)$ unintegrated (since we have not yet
introduced its corresponding antighost), but we can now integrate
over both $\psi_\mu(x)$ and $A^*_\mu(x)$. This leads to the following
extended, but not yet fully gauge-fixed, action $S_{ext}$:
\begin{eqnarray}
{\cal{Z}} &=& \int [dA_\mu][da_\mu][d\psi_\mu][dA^*_\mu][db_\mu]
\exp\left[\frac{i}{\hbar}S_{ext}\right] \cr
S_{ext} &=& S[A_\mu-a_\mu] - \int dx\{b_\mu(x)a_\mu(x) +
A^*_\mu(x)[\psi^\mu(x) - D_{(A-a)}^\mu c(x)]\}
\end{eqnarray}
This extended action is invariant under the BRST transformation (38).
The full integration measure is also invariant. Of course, the expression
above is still formal, since we have not yet gauge fixed ordinary
Yang-Mills invariance. (This is obvious from our construction, but can
also be checked explicitly. The easiest way is to first integrate out
$a_\mu$ and $b_\mu$; integration over $A^*_\mu$ then finally leaves a
trivial $\psi_\mu$-integral. What is left over is nothing but the
starting point, the Yang-Mills action $S[A_\mu]$ integrated over the
original measure.)

Furthermore, we are eventually going to integrate over the ghost
$c$, which already now appears in the extended action. If we insist
that the Schwinger-Dyson equations involving this field, $i.e$,
equations of the form
\beq
0 = \int [dc] \frac{\delta^l}{\delta c(x)}\left[F e^{\frac{i}{\hbar}
\left[\mbox{\rm Action}\right]}\right]
\eeq
are to be satisfied automatically (for reasonable choices of functionals
$F$) by means of the full unbroken BRST
algebra, we must introduce yet one more collective field. This
new collective field, call it $\tilde{c}(x)$, is Grassmann odd, and
has $gh(\tilde{c}) = 1$. Now shift the Yang-Mills ghost:
\beq
c(x) \to c(x) - \tilde{c}(x) .
\eeq
To fix the associated fermionic gauge symmetry, we introduce a new
BRST multiplet of a ghost-antighost pair and an auxiliary field. We
follow the same rule as before and let the transformation of the new
collective field $\tilde{c}$ carry the (BRST) transformation of the
original ghost $c$, $viz.$,
\begin{eqnarray}
\delta c(x) &=& C(x) \cr
\delta \tilde{c}(x) &=& C(x) + \frac{1}{2}[c(x)-\tilde{c}(x),
c(x)-\tilde{c}(x)] \cr
\delta C(x) &=& 0 \cr
\delta c^*(x) &=& B(x) \cr
\delta B(x) &=& 0 .
\end{eqnarray}

It follows from (43) that ghost number assignments should be as
follows:
\beq
gh(C) = 2~,~~~ gh(c^*) = -2~,~~~ gh(B) = -1 ,
\eeq
and $B$ is a fermionic Nakanishi-Lautrup field. Fields of the kind
$C$ and $c^*$ are sometimes known as ghosts for ghosts. However, in
contrast to conventional examples involving such ghosts, these are
not a priori required in this context. They enter only when we by
hand gauge the ghost-shift symmetry.

Next, gauge-fix $\tilde{c}(x)$ to zero by adding a term
\beq
-\delta[c^*(x)\tilde{c}(x)] =  B(x)\tilde{c}(x) -
c^*(x)\{C(x) + \frac{1}{2}[c(x)-\tilde{c}(x),c(x)-\tilde{c}(x)]\}
\eeq
to the Lagrangian. This gives us the fully extended action,
\begin{eqnarray}
S_{ext} &=& S[A_\mu-a_\mu] - \int dx\{b_\mu(x)a_\mu(x) +
A^*_\mu(x)[\psi^\mu(x) - D^\mu_{(A-a)}\{c(x)- \tilde{c}(x)\}] \cr
&& - B(x)\tilde{c}(x) + c^*(x)(C(x)+\frac{1}{2}
[c(x)-\tilde{c}(x),c(x)-\tilde{c}(x)])\} ,
\end{eqnarray}
with the partition function so far being integrated over all fields
appearing above, except for $c$ (whose antighost $\bar{c}$ still has
to be introduced when we gauge-fix the original Yang-Mills symmetry).

The extended action {\em and} the functional measure is formally
invariant under the following set of transformations:\footnote{It may
be necessary to comment on the invariance of the measure.
Since a path integral measure needs regularization, a statement about
invariance or non-invariance under the BRST symmetry
really requires a more careful analysis. But there is a difference
between a formally invariant
measure and a formally non-invariant one: A formally invariant
measure remains invariant within a regularization scheme that does
not break the symmetry, while very specific regularization schemes
are required to enforce invariance of a formally non-invariant measure.
To give a more precise meaning to the difference between formally
invariant or non-invariant measures one can consider the case of a
finite number of degrees of freedom, where no regularization is needed.
Dimensional regularization is believed to leave, in cases where it is
meaningful, all measures
invariant under local transformations since such
transformations normally lead to factors of $\delta(0)$. At the
level of Feynman diagrams, such factors are equated with zero in
dimensional regularization. Of course, if for one reason or another
one insists on using a regularization scheme that explicitly breaks
the above BRST symmetry, then the measure -- so defined -- may be
non-invariant even in the limit of removing the regulator. One
should then try to add appropriate local counterterms to the action,
counterterms whose BRST variations should precisely compensate for the
non-invariance of the measure. This situation is no different from
what happens in standard gauge-fixed Yang-Mills theory if one insists
on regulators that break the usual Yang-Mills BRST symmetry. For
a recent discussion of this, of true anomalies, and one-loop
counterterms within the Batalin-Vilkovisky framework
see ref. \cite{Troost}.}

\vspace{0.6cm}
$
\begin{array}{llllll}
\delta A_\mu(x) &=& \psi_\mu(x) , & \delta \psi_\mu(x)
&=& 0 , \\
\delta a_\mu(x) &=& \psi_\mu(x)-D^{(A-a)}_\mu[c(x)-\tilde{c}(x)]
, & \delta c(x)
&=&  C(x) ,\\
\delta A^*_\mu(x) &=& b_\mu(x) & \delta b_\mu(x) , &=& 0 ,\\
\delta \tilde{c}(x) , &=& C(x) +
\frac{1}{2}[c(x)-\tilde{c}(x),c(x)-\tilde{c}(x)]  &
\delta C(x) &=& 0, \\
\delta c^*(x) &=& B(x) & \delta B(x) &=& 0 ~.
\end{array}
$
\vspace{0.6cm}

As the notation indicates, the fields $A^*_\mu(x)$ and $c^*(x)$ can be
identified with the Batalin-Vilkovisky antifields of $A_\mu(x)$ and
$c(x)$, respectively. {\em These antifields are the usual antighosts
of the collective fields enforcing Schwinger-Dyson equations through
shift symmetries.}

Note that the general rule of assigning ghost number and Grassmann
parity to the antifields,
\beq
gh(\phi^*_A) = - (gh(\phi^A) + 1)~,~~~~~ \epsilon(\phi^*_A) =
\epsilon(\phi^A) + 1~,
\eeq
arises in a completely straightforward manner. Here, it is a
simple consequence of the fact that the BRST operator raises ghost
number by one unit (and changes statistics), supplemented with the
usual rule that antighosts have opposite ghost number of the ghosts.

The extended action (46) above has more fields than the extended
action of Batalin and Vilkovisky, and the transformation laws of what
we identify as antifields do not match those of ref. \cite{Batalin}.
In the form we have presented it, the BRST symmetry is nilpotent also
off-shell. If we integrate over the auxiliary fields $b_\mu$ and
$B$ (and subsequently over $a_\mu$ and $\tilde{c}$) the BRST symmetry
becomes nilpotent only on-shell. The extended action then takes the
following form:\footnote{We are dropping the subscript
on the covariant derivative since no confusion can arise at this
point.}
\beq
S_{ext} = S[A_\mu] - \int dx\{A^*_\mu(x)[\psi^\mu(x) -
D^\mu c(x)] + c^*(x)(C(x)+\frac{1}{2}[c(x),c(x)])\} .
\eeq
This differs from the extended action of Batalin and Vilkovisky by
the terms involving the ghost fields $\psi_\mu(x)$ and $c^*(x)$.
Comparing with the
case of no gauge symmetries, this is exactly what we should expect.
These ghost fields $\psi_\mu$ and $c^*$ ensure the correct
Schwinger-Dyson equations for $A_\mu$ and $c$, respectively.

To find the corresponding BRST symmetry we use -- as justified earlier --
the equations of motion for the auxiliary fields $b_\mu$ and $B$, and
use the $\delta$-function constraints on $a_\mu$ and $\tilde{c}$. This
gives
\begin{eqnarray}
\delta A_\mu(x) &=& \psi_\mu(x) \cr
\delta\psi_\mu(x) &=& 0 \cr
\delta c(x) &=& C(x) \cr
\delta C(x) &=& 0 \cr
\delta A^*_\mu(x) &=& - \frac{\delta^l S_{ext}}{\delta A_\mu(x)} \cr
\delta c^*(x) &=& - \frac{\delta^l S_{ext}}{\delta c(x)}
\end{eqnarray}

BRST invariance of the extended action (48) immediately implies
that it satisfies a Master Equation we can write as
\beq
\int dx\frac{\delta^r S_{ext}}{\delta A^*_{\mu}(x)}
\frac{\delta^l S_{ext}}{\delta A^{\mu}(x)} + \int dx
\frac{\delta^r S_{ext}}{\delta c^*(x)}
\frac{\delta^l S_{ext}}{\delta c(x)}
= \int dx\frac{\delta^r S_{ext}}{\delta A_{\mu}(x)}\psi_{\mu}(x) +
\int dx\frac{\delta^r S_{ext}}{\delta c(x)}C(x) ~.
\eeq

Note that this Master Equation precisely is of the form
\beq
\frac{1}{2} (S_{ext},S_{ext}) = - \int dx \frac{\delta S_{ext}}
{\delta \phi^A(x)}c^A(x) ~,
\eeq
with two ghost fields $c^A$ that are just $\psi_{\mu}$ and
$C(x)$.

To finally make contact with the Batalin-Vilkovisky formalism, let us
integrate out the ghost $\psi_\mu$. As in section 2, we shall use an
identity of the form
\begin{eqnarray}
&&\int[d\psi_\mu]F[\psi^\mu(y)]\exp\left[-\frac{i}{\hbar}\int dx
A^*_\mu(x)\{\psi^\mu(x) - D^\mu c(x)\}\right] \cr
&=&F\left[D^\mu c(y)+ (i\hbar)\frac{\delta^l}{\delta A^*_\mu(y)}\right]
\int [d\psi_\mu] \exp\left[-\frac{i}{\hbar}\int dx
A^*_\mu(x)\{\psi^\mu(x) - D^\mu c(x)\}\right]  \cr
&=&\exp\left[\frac{i}{\hbar}\int dx A^*_{\mu}(x)D^{\mu}c(x)\right]
F\left[(i\hbar)\frac{\delta^l}{\delta A^*_{\mu}(y)}\right]
\delta(A^*_{\mu})
\end{eqnarray}

This shows that we should replace $\psi_\mu(x)$ by its equation of
motion, plus the shown quantum correction of $\cal{O}(\hbar)$
which then acts on the rest of the integral, or equivalently, by
just the derivative operator (which then acts solely on the
functional $\delta$-function). To get a useful representation of
$\psi^{\mu}(y)$ we integrate the latter version of the identity
by parts, thus letting the derivative
operator act on everything except $\delta(A^*_{\mu})$. This
automatically brings down the equations of motion for $\psi^{\mu}$.
(An equivalent series of steps can of course be made using the
former version of the identity, and the result is the same).

Having in this manner integrated out $\psi^{\mu}$ and $C$, the
partition function reads
\begin{eqnarray}
{\cal{Z}} &=& \int [dA_\mu][dA^*_\mu][dc^*]\delta(A^*_\mu)
\delta(c^*)\exp\left[\frac{i}{\hbar}S_{ext}\right] \cr
S_{ext} &=& S[A_\mu] + \int dx\{A^*_\mu(x)D^\mu c(x) -
\frac{1}{2}c^*[c(x),c(x)]\} ,
\end{eqnarray}
and the {\em classical} BRST symmetry follows as discussed above
by substituting only the equations of motion for $\psi_\mu$ and
$C$:
\begin{eqnarray}
\delta A_\mu(x) &=& \frac{\delta^l S_{ext}}{\delta A^*_{\mu}(x)}
= D_\mu c(x) \cr
\delta c(x) &=& \frac{\delta^l S_{ext}}{\delta c^*(x)}
-\frac{1}{2}[c(x),c(x)] \cr
\delta A^*_\mu(x) &=& -\frac{\delta^l S_{ext}}{\delta A_\mu(x)}
\cr \delta c^*(x) &=& -\frac{\delta^l S_{ext}}{\delta c(x)} ~.
\end{eqnarray}

This is the usual extended action of Batalin and Vilkovisky and the
corresponding classical BRST symmetry. Of course, in the partition
function the integrals over $A^*_\mu$ and $c^*$ are trivial. This
is as it should be, because by integrating out these antighosts we
should finally recover the starting measure and the still not
gauge-fixed Yang-Mills action. What we have provided here is thus
only a very precise functional derivation of the extended action. It
shows that we {\em can} understand the extended action in the usual
path integral framework, and that the integration measures for
$A^*_\mu$ and $c^*$ (with the accompagnying $\delta$-function
constraints) are provided automatically.

Precisely because of this prescribed path integral measure for
what are really the antifields $A^*_\mu$ and
$c^*$, it may appear as if this extended action does not have much
independent significance: All additional terms in the action
actually vanish because of the $\delta$-function constraints.
Related to this is the fact that we could have derived an identity
equivalent to eq. (52) by first shifting
\beq
\psi_\mu(x) \to \psi_\mu(x) + D_\mu c(x)
\eeq
(and the right hand side of eq. (52) would in that case not contain
the $D_\mu c(x)$ part in the exponential). Similarly, $C$ could be
shifted as
\beq
C(x) \to C(x) -\frac{1}{2}[c(x),c(x)] ~,
\eeq
and the extended action would then contain neither $A^*_{\mu}$ nor
$c^*$.
At the same time the BRST transformations (49) would
be changed, giving again the equations of motion for the shift
ghosts $\psi^{\mu}$ and $C$ (which just corresponds to the usual
Yang-Mills BRST symmetry). This ambiguity in the
derivation of the extended action is directly related to the fact that
the split of symmetries as in eq. (35) is arbitrary. The split given
in eq. (35) is what leads to the extended action of Batalin and
Vilkovisky; the prescription is to choose the
collective fields to carry the original gauge (or BRST) symmetry in
their transformations. There is a one-to-one correspondence between
this ambiguity and the freedom to choose boundary conditions when
solving the Batalin-Vilkovisky Master Equation.

There should be no special significance attached to the way
we shuffle the internal symmetry transformations between the
original fields and the collective fields. Since there is a
direct correspondence between this and the choice of boundary
conditions in the Batalin-Vilkovisky formalism, it must mean
that these boundary conditions can be changed. This is indeed
the case (as will be shown in the next section), but the cost
one pays is that gauge fixing of the internal symmetries may then
no longer correspond to canonical transformations within the
antibracket (see below).

Another very important remark is
the following. Although the classical BRST symmetry (54) is a
symmetry of the action, it is simply not a correct symmetry of the
theory because the functional measure is not invariant due to
the presence of $\delta$-function constraints on the antifields. So
this classical BRST symmetry is of very limited use. In particular,
if we derive standard BRST Ward Identities based on eq. (54) we
find that they are correct only up to ${\cal{O}}(\hbar)$. In terms of
Schwinger-Dyson equations, these Ward identities correspond only
to the trivial classical part ($i.e.$ terms involving
the equations of motion). Since it is not useful to do
classical field theory this way, the split into a classical and
a quantum part of the BRST symmetry is rather unfortunate. The
{\em full} BRST symmetry of the extended action (53) follows from
our derivation above:
\begin{eqnarray}
\delta A^\mu(x) &=& D^\mu c(x)
+ i\hbar\frac{\delta^r}{\delta A^*_\mu(x)}
\cr
\delta c(x) &=& -\frac{1}{2}[c(x),c(x)] - i\hbar\frac{\delta^r}
{\delta c^*(x)} \cr
\delta A^*_\mu(x) &=& -\frac{\delta^l S_{ext}}{\delta A_\mu(x)} \cr
\delta c^*(x) &=& -\frac{\delta^l S_{ext}}{\delta c(x)} ~.
\end{eqnarray}

Since this set of transformations involves operators, an interpretation
is required. Namely, the replacement of ghost fields by equations of
motion plus a quantum correction is only valid inside the path integral.
It is indeed a symmetry
of the combination of action and measure. As a check, the Ward
Identities derived from this symmetry are the correct Schwinger-Dyson
equations.

Let us stress once again that the quantum BRST symmetry (57) from the
present point of view is an awkward but of course {\em bona
fide} representation of the combined Yang-Mills and
Schwinger-Dyson BRST algebra. It is obtained only on the insistence
of integrating out ghost fields while keeping their corresponding
antighosts. The original BRST symmetry (49) automatically includes
classical and quantum effects simultaneously, as is usual for
internal symmetries.

When we finally (after Yang-Mills gauge fixing, which we shall turn
to shortly) also integrate over $c^*$ and $A^*_{\mu}$, the full
left-over BRST transformations equal the usual Yang-Mills BRST
transformations.

Obviously, the extended action is not yet very useful from the point
of view of ordinary BRST gauge fixing. To unravel some of the
mechanisms behind the Batalin-Vilkovisky scheme, it is nevertheless
advantageous to keep the -- at this point somewhat superfluous
-- antifields.
In fact, it is even more useful to return to the formulation in eq.
(48), where there is yet no split into a classical and a quantum part
of the symmetry. Let us therefore take (48) as the starting action,
and now just gauge-fix in a standard manner the Yang-Mills symmetry.
Choosing, $e.g.$, a covariant gauge, we therefore finally extend the
BRST multiplet to include a Yang-Mills antighost $\bar{c}$ and a
Nakanishi-Lautrup scalar $b$. For there to be no doubt, let us also
note that these fields have
\beq
gh(\bar{c}) = -1~,~~~~ gh(b) = 0 ~.
\eeq
The BRST transformations are the usual $\delta \bar{c}(x) = b(x)$
and $\delta b(x) = 0$.
Gauge fixing to a covariant $\alpha$-gauge can be achieved by adding
a term
\beq
\delta[\bar{c}(x)\{\partial_\mu A^\mu(x) - \frac{1}{2\alpha}b(x)\}] =
b(x)\partial_\mu A^\mu(x) + \bar{c}(x)\partial_\mu\psi^\mu(x)
- \frac{1}{2\alpha}b(x)^2
\eeq
to the Lagrangian. The corresponding completely gauge-fixed extended
action then reads
\begin{eqnarray}
S_{ext} &=& S[A_\mu] - \int dx\{A^*_\mu(x)[\psi^\mu(x) -
D^\mu c(x)] + c^*(x)(C(x)+\frac{1}{2}[c(x),c(x)]) \cr
&& - b(x)\partial_\mu
A^\mu(x) - \bar{c}(x)\partial_\mu\psi^\mu(x)
+ \frac{1}{2\alpha}b(x)^2\} .
\end{eqnarray}

Now integrate out $\psi^\mu$ and $C$. The result is indeed a
partition function of the form suggested by the formalism of
Batalin and Vilkovisky:
\begin{eqnarray}
{\cal{Z}} &=& \int [dA_\mu][dA^*_\mu][d\bar{c}][dc][dc^*][db]
\delta(A^*_\mu + \partial_\mu\bar{c})\delta(c^*)e^{\left[
\frac{i}{\hbar}S_{ext}\right]} \cr
S_{ext} &=& S[A_\mu] + \int dx\{A^*_\mu(x)D^\mu c(x)
- \frac{1}{2}c^*(x)[c(x),c(x)] \cr
&& +b(x)\partial_\mu A^\mu(x)
- \frac{1}{2\alpha}b(x)^2\} .
\end{eqnarray}

Note that by adding the Yang-Mills gauge-fixing terms, the
$\delta$-function constraint on the antifield $A^*_\mu$ has been shifted.
Thus when doing the $A^*_\mu$-integral, we are in effect substituting
not $A^*_\mu(x) = 0$ but
\beq
A^*_\mu(x) = -\partial_{\mu}\bar{c}(x) =
\frac{\delta^r \Psi}{\delta A^\mu(x)} ~,
\eeq
where $\Psi$ is defined as the term whose BRST variation is added to the
action, $i.e.$, in this particular case,
\beq
\Psi = \int dx\{\bar{c}(x)(\partial^\mu A_\mu(x) - \frac{1}{2\alpha}
b(x))\} ~.
\eeq
Upon doing the $A^*_\mu$ and $c^*$ integrals, one recovers the
standard covariantly gauge-fixed Yang-Mills theory
\beq
S = S[A_\mu] + \int dx\{\bar{c}(x)\partial^\mu D_\mu c(x) +
b(x)\partial^\mu A_\mu(x) - \frac{1}{2\alpha}b(x)^2\}.
\eeq

The identification (62) strongly suggests that one can see gauge
fixing as a particular canonical transformation involving new fields
(the antighosts $\bar{c}$). But there are terms in eq. (64) (those
involving $b(x)$) which
do not immediately follow from this perspective. In the
Batalin-Vilkovisky
framework, this is resolved by noting that one can always add
terms of new fields and antifields with trivial antibrackets.
In this version of the gauge-fixing procedure, one returns
to the (``minimally") extended action of eq. (61)
and extends it in a ``non-minimal" way.  In
the Yang-Mills case, this includes an additional term in the
action of the form
\beq
S_{nm} = \int dx \bar{c}^*(x)b(x) ~,
\eeq
with $\bar{c}^*$ and $b$ having the same ghost number and Grassmann
parity:
\beq
gh(\bar{c}^*) = gh(b) = 0 ~;~~~~~~\epsilon(\bar{c}^*) =
\epsilon(b) = 0 ~.
\eeq
As the notation indicates, these new fields $\bar{c}^*$ and $b$ are
indeed just the antifield of $\bar{c}$, and the usual
Nakanishi-Lautrup field, respectively.

Gauge fixing to the same gauge as above can then be achieved by
the same gauge fermion (63) which now affects both $A^*_{\mu}$ and
$\bar{c}^*$ within the antibracket.
It can thus be seen as the canonical transformation
that shifts $A^*_{\mu}$ and $\bar{c}^*$ from zero to
\beq
A^*_{\mu}(x) = \frac{\delta^r \Psi}{\delta A^{\mu}(x)} ~,~~~~~~
\bar{c}^*(x) = \frac{\delta^r \Psi}{\delta \bar{c}(x)} ~.
\eeq
Since $\Psi$ does not depend on the antifields, this canonical
transformation leaves all {\em fields} $A_{\mu}, c$ and $\bar{c}$
unchanged.

Can we understand the non-minimally extended action from our point
of view too? Consider the stage at which we introduce the antighost
$\bar{c}$. This field does not yet appear in the action, but we can
of course still introduce a corresponding collective ``shift" field
$\bar{c}'$ for $\bar{c}$
as well.
The corresponding BRST multiplet consists of a new ``shift-antighost"
$\lambda(x)$, an ``anti-antighost" $\bar{c}^*(x)$, and the associated
auxiliary field $B'(x)$:
\begin{eqnarray}
\delta \bar{c}(x) &=& \lambda(x) \cr
\delta \bar{c}'(x) &=& \lambda(x) - b(x) \cr
\delta \lambda(x) &=& 0 \cr
\delta b(x) &=& 0 \cr
\delta \bar{c}^*(x) &=& B'(x) \cr
\delta B'(x) &=& 0 ~.
\end{eqnarray}
The assignments will then have to be exactly as in eq. (66),
supplemented with $gh(B') = \epsilon(B') = 1$. We are again dealing
with {\em two} symmetries, because the shifted field $\bar{c}(x) -
\bar{c}'(x)$ itself can still be shifted by the usual Nakanishi-Lautrup
field. Let us now gauge-fix this huge symmetry. We do
it in the most simple manner by adding a term
\beq
-\delta[\bar{c}^*(x)\bar{c}'(x)] = B'(x)\bar{c}'(x) - \bar{c}^*(x)(
\lambda(x) - b(x))
\eeq
to the Lagrangian. The integrals over $B'$ and $\bar{c}'$ are of course
trivial and we are left with the non-minimally extended
action for this theory plus, as expected, the corresponding term with
the new ghost $\lambda$. The final gauge-fixing of the Yang-Mills
symmetry will now consist in adding, instead of eq. (59),
\beq
\delta[\bar{c}(x)\{\partial_\mu A^\mu(x) - \frac{1}{2\alpha}
b(x)\}] =
\lambda(x)\partial_\mu A^\mu(x) + \bar{c}(x)\partial_\mu\psi^\mu(x)
- \frac{1}{2\alpha}\lambda(x)b(x) ~.
\eeq

Before Yang-Mills gauge fixing, the integral over $\lambda(x)$ just gave
a factor of $\delta(\bar{c}^*)$. After adding the gauge-fixing term,
this is changed:
\beq
\delta(\bar{c}^*) \to \delta(\bar{c}^*(x) - \partial^{\mu}A_{\mu}(x) +
\frac{1}{2\alpha}b(x))~.
\eeq
Substituting this back into extended action, we recover the result (64).
Note that this indeed can be viewed as a canonical transformation within
the antibracket. All the correct $\delta$-function constraints are
provided by the collective fields and their ghosts. Since the functional
$\Psi$ has been chosen to depend only on the fundamental fields, and
not on the antifields, the fields $A_{\mu}, c$ and $\bar{c}$ are all
left untouched by this canonical transformation. Extending the action
from the minimal to the non-minimal case is equivalent to demanding that
also Schwinger-Dyson equations for $\bar{c}(x)$ follow as Ward identities
of the BRST symmetry. Since the antighost $\bar{c}$ remains in the path
integral after gauge fixing, it would indeed be very unnatural not to
demand that correct Schwinger-Dyson equations for this field follow as
well. As shown, this requirement automatically leads to the
{\em non-minimally} extended action.

As for the functional measure, we have stressed earlier that
we always assume the existence of a suitable regulator that
preserves the pertinent BRST symmetry. We can make this statement
a little more explicit by detailing the required symmetries of
the measure in this Yang-Mills case. Before integrating out any
fields, the measures for $A_{\mu}, c$ and $\bar{c}$ should all
be invariant under local shifts. For $A_{\mu}$ this corresponds
to the usual euclidean measure (and it is very difficult to imagine
this shift symmetry being broken by any reasonable regulator),
while for $c$ and $\bar{c}$ this is consistent with the usual rules
of Berezin integration. The measures for the three collective
fields should in addition be invariant under what corresponds to
usual Yang-Mills BRST transformations, a property that indeed holds
formally. Finally, the measures for all antifields are only
required to be invariant under local shifts. After having
integrated out the auxiliary fields $B_A$, invariance of these
measures of the antifields is now non-trivial but one can check
explicitly that it is formally satisfied. This should indeed
be the case, because it is straightforward to check that the
action remains invariant. Since at least the {\em combination}
of measure and action must remain invariant after integrating
out some of the fields, invariance of the measure is in this
case formally guaranteed.

Let us finally point out that once the antighost $\bar{c}$ is being
treated on equal footing with $A_{\mu}$ and $c$, a Master Equation
of the form
\beq
\frac{1}{2}(S_{ext},S_{ext}) = - \int dx\frac{\delta^r S_{ext}}
{\delta\phi^A(x)}c^A(x)
\eeq
now holds with $\phi^A$ denoting {\em all} the fields that finally
remain in the path integral: $A_{\mu},c$ and $\bar{c}$. Similarly,
the BRST algebra becomes, upon integrating out the collective fields
$\varphi^A$ and the auxiliary fields $B_A$, of the very simple form
(21) we encountered already in the case of no internal gauge
symmetries.

\section{Generalizations}

The two previous sections were useful for illuminating in familiar
settings the way the Batalin-Vilkovisky formalism arises from more
standard quantization principles. The main new requirement is that
the internal symmetry algebra should contain what we call the
Schwinger-Dyson BRST symmetry. At a formal level, this guarantees
that the full quantum theory is entirely determined by the classical
action and the complete BRST symmetry.

At this stage it may be useful to step back and extract the basic
ingredients of the analysis in a more condensed and general manner.
As should be clear already from the Yang-Mills case, just the
insistence on incorporating the Schwinger-Dyson BRST symmetry is
not in itself sufficient to guarantee that the canonical formalism
of Batalin and Vilkovisky follows. The part of possible
representations of gauge theories that fall into this canonical
framework is small, and we have to tune
carefully the prescription in order to regain the Batalin-Vilkovisky
formalism upon integrating out certain ghost fields. One possible
advantage of this fact is that, although we can reproduce known results
based on the antifield-antibracket formalism, we know now that
there is ample scope for generalizations.

Having included all space-time integrations in a detailed manner in
the two previous sections, we shall here for convenience of notation
drop these integrations, and, as in section 1, consider the summation
convention of repeated indices to include space-time summations as
well.
Consider first a theory which is invariant under an irreducible set of
gauge transformations that form a closed algebra:
\beq
\delta \phi^A = R^A_{\alpha} \epsilon^{\alpha} ~.
\eeq
Both theories without gauge symmetries and, $e.g.$, Yang-Mills theory
fall into this category. The general procedure is now the following.
Introduce collective fields such that Schwinger-Dyson equations
for all fundamental fields ($i.e.$, all the fields that will
eventually appear in the gauge-fixed action) are satisfied as Ward
Identities of a BRST symmetry. To simplify the notation, let us group
all fundamental fields into the same $\phi^A$. In, $e.g.$ the
Yang-Mills case this field $\phi^A$ will then contain both $A_{\mu}, c$
and $\bar{c}$. When the functional measures are flat,
enforcing Schwinger-Dyson equations can be done through shifts:
\beq
\phi^A \to \phi^A - \varphi^A
\eeq
The action is now invariant under arbitrary deformations of the
fundamental fields, all those fields that remain after the gauge
fixing. The next step is to form the nilpotent BRST algebra that
incorporates these shifts. To get to the antibracket formalism, one
should let the fundamental fields transform just as those shifts,
$i.e.$, $\delta \phi^A = c^A$. (The ghosts $c^A$ are now shift ghosts;
they should not be confused with, $e.g.$, the usual Yang-Mills ghosts.)
Furthermore, one should include the usual gauge (and BRST) symmetries
in the variations of the collective fields. As we shall see shortly,
these requirements are equivalent to certain boundary conditions in
Batalin-Vilkovisky formalism. To complete the BRST multiplet one
finally introduces the corresponding antighosts, here denoted by
$\phi^*_A$. The full nilpotent BRST transformations are then
\begin{eqnarray}
\delta \phi^A &=& c^A \cr
\delta \varphi^A &=& c^A - {\cal{R}}^A[\phi-\varphi] \cr
\delta c^A &=& 0 \cr
\delta b^A &=& 0 \cr
\delta \phi^*_A &=& B_A \cr
\delta B_A &=& 0~,
\end{eqnarray}
where $\cal{R}^A$ is the BRST generalization of the gauge generator of
eq. (73) to the full set of both usual fields, usual ghosts and
usual antighosts. It fulfills $\delta{\cal{R}}^A = 0$.

The gauge-fixing procedure can now be done in one step. One fixes all
collective fields to zero, and introduces a ``gauge fermion" $\Psi$
to fix the underlying gauge symmetry (if there is any). This is done
by adding a term
\beq
-\delta[\phi^*_A\varphi^A - \Psi[\phi]] = (-1)^{\epsilon_A+1}
B_A\varphi^A - \phi^*_A(c^A - {\cal{R}}^A[\phi-\varphi]) +
\frac{\delta^r \Psi}{\delta\phi^A}c^A
\eeq
to the action, which then reads
\beq
S_{g.f.} = S[\phi-\varphi] + (-1)^{\epsilon_A+1}B_A\varphi^A
- \phi^*_A(c^A - {\cal{R}}^A[\phi-\varphi]) +
\frac{\delta^r \Psi}{\delta\phi^A}c^A ~.
\eeq

The final gauge-fixed partition function is thus of the simple form
\beq
{\cal{Z}} = \int [d\phi][d\varphi][d\phi^*][dB] \exp\left[
\frac{i}{\hbar} S_{g.f.}\right] ~,
\eeq
and both the gauge-fixed action and the functional measure are
formally invariant under the nilpotent BRST symmetry above. The
nilpotency holds also off-shell. The functional measures for
$\phi^A$ and $\phi^*_A$ are formally invariant since they are
presumed flat. The measure for $\varphi^A$ should in addition be
invariant under the transformations ${\cal{R}}^A$.

Integrating out $B_A, \varphi^A$ and $c^A$ leads to
\begin{eqnarray}
{\cal{Z}} &=& \int [d\phi][d\phi^*] \delta\left(\phi^*_A -
\frac{\delta \Psi}{\delta^r \phi^A}\right)\exp\left[\frac{i}{\hbar}
S_{g.f.}\right] \cr
S_{g.f.} &=& S[\phi] + \phi^*_A{\cal{R}}^A ~.
\end{eqnarray}
This coincides with the Batalin-Vilkovisky gauge-fixed action for
this case. We see that the final replacement of the ``antifields"
$\phi^*_A$ with $\delta \Psi/\delta\phi^A$ is a direct consequence
of having integrated out the shift-ghosts $c^A$.

To see how the antibracket formalism emerges in this more general
setting, split up the gauge-fixing procedure in two steps. First
gauge-fix the collective fields to zero, without gauge-fixing the
underlying gauge symmetry. This is achieved by adding only
\beq
-\delta[\phi^*_A\varphi^A] = (-1)^{\epsilon_A+1}B_A\varphi^A
- \phi^*_A(c^A - {\cal{R}}^A[\phi-\varphi])
\eeq
to the action, which then reads
\beq
S_{ext} = S[\phi-\varphi] + (-1)^{\epsilon_A+1}B_A\varphi^A
- \phi^*_A(c^A - {\cal{R}}^A[\phi-\varphi]) ~.
\eeq

Now consider integrating out the auxiliary fields $B_A$
and the collective fields $\varphi^A$ first. The extended action
then becomes
\beq
S_{ext} = S[\phi] - \phi^*_A(c^A - {\cal{R}}^A[\phi]) ~.
\eeq

At the same time, this changes the BRST algebra into
\begin{eqnarray}
\delta \phi^A &=& c^A \cr
\delta c^A &=& 0 \cr
\delta \phi^*_A &=& - \frac{\delta^l S_{ext}}
{\delta \phi^A} ~.
\end{eqnarray}

Demanding that the extended action $S_{ext}$ is invariant under this
BRST symmetry is then equivalent to
\beq
0 = \delta S_{ext} = \frac{\delta^r S_{ext}}{\delta\phi^A} c^A -
\frac{\delta^r S_{ext}}{\delta \phi^*_A}\frac{
\delta^l S_{ext}}{\delta \phi^A} ~,
\eeq
or, in other words, precisely a Master Equation of the general form
\beq
\frac{1}{2}(S_{ext},S_{ext}) = -
\frac{\delta^r S_{ext}}{\delta\phi^A}c^A~.
\eeq

For irreducible closed gauge algebras this is the end of the story.
The Master Equation (85) contains no terms of order $\hbar$ or higher.
It is a classical equation that can be solved algebraically, without
resort to $\hbar$-expansions. For theories with ultraviolet divergences
this holds as long as the chosen regularization scheme respects the
BRST symmetry, but the existence of such a regularization procedure
has been the working assumption throughout. We refer again to the
footnote in section 3 concerning this issue.

If one finally gauge-fixes the underlying gauge symmetry, one adds
a term
\beq
\delta \Psi[\phi] =  \frac{\delta^r\Psi}{\delta\phi^A}c^A
\eeq
to the action. The gauge-fixed partition function is then
again of the form (79).

It is important that the ``gauge fermion" $\Psi$ is a function of the
fundamental fields only. This is because the BRST Schwinger-Dyson
symmetry (83) is not nilpotent in general. A term of the form
$\delta\Psi$ is therefore not automatically BRST invariant. It will
of course have vanishing expectation value, but the exponentiated
form $\exp[\delta\Psi]$ will not in general, and its presence will
affect Green functions. However,
the operator $\delta$ of eq. (83) {\em is} nilpotent when acting
on the fields only: $\delta^2\phi^A = 0$. We are therefore permitted
to add a (then BRST-exact) term of the form $\delta\Psi$, with $\Psi$
being a function of just the fields $\phi^A$. Adding this term will
for reasonable choices of $\Psi$ not affect BRST-invariant expectation
values. This means also that the appropriate Schwinger-Dyson
equations are preserved in this procedure.

Because of the symmetry properties of the antibracket with respect
to bosonic objects such as the action $S_{ext}$, we happened to
extract the antibracket in the Master Equation (85) even though we
in effect have only half of the canonical structure upon integrating
out the collective fields $\varphi^A$ and the auxiliary fields
$B_A$. Integrating out also the shift
ghosts $c^A$ while keeping the antighosts $\phi^*_A$ yields the
missing ingredients of the full canonical structure behind the
antibracket. The quantum corrections to the BRST symmetry which we
discussed in detail in the two previous sections can, however, not
be ignored. They distort the canonical structure by the quantum
operator
\beq
i\hbar \Delta = (-1)^{\epsilon_A+1}(i\hbar)\frac{\delta^r\delta^r}
{\delta\phi^*_A\delta\phi^A}~.
\eeq
We shall discuss this in some more detail in section 4.2.

The addition of the term from the ``gauge fermion" $\Psi$ to the
action can, as we have seen above, be viewed as a canonical
transformation from $\phi^*_A = 0$ to $\phi^*_A = \delta^r\Psi/
\delta\phi^A$. It is not an arbitrary canonical transformation
however, because $\Psi$ must in general be a function of the
fields $\phi^A$ only. The reason for this restriction is very
clear in the present formulation, because the BRST operator is
only nilpotent on the $\phi^*_A$-independent subspace.

In the case of an irreducible closed gauge algebra, the boundary
conditions
one imposes on the Master Equation have direct counterparts in the
collective field formalism. Recall that we always have the freedom to
shuffle internal gauge symmetries between the fundamental fields and
the collective fields. Choosing to let the transformations of the
fundamental fields be {\em only} arbitrary shifts (thereby lumping
all internal transformations into the transformations of the collective
fields) is equivalent to
specifymusting the boundary conditions of Batalin and Vilkovisky.

The freedom in specifying the boundary conditions can be made quite
manifest by means of another choice of transformations for the
fundamental fields and the collective fields. Suppose that instead
of eq. (75) we choose to let the collective fields transform only
as shifts. The BRST symmetry is then
\begin{eqnarray}
\delta \phi^A &=& c^A + {\cal{R}}^A[\phi] \cr
\delta \varphi^A &=& c^A \cr
\delta c^A &=& 0 \cr
\delta b^A &=& 0 \cr
\delta \phi^*_A &=& B_A \cr
\delta B_A &=& 0~,
\end{eqnarray}
instead of (75). Gauge fixing the collective fields to zero is now
achieved by adding
\beq
-\delta[\phi^*_A\varphi^A - \Psi[\phi]] = (-1)^{\epsilon_A+1}
B_A\varphi^A - \phi^*_Ac^A +
\frac{\delta^r \Psi}{\delta\phi^A}(c^A + {\cal{R}}^A[\phi])
\eeq
to the action, which then reads
\beq
S_{g.f.} = S[\phi-\varphi] + (-1)^{\epsilon_A+1}B_A\varphi^A
- \phi^*_Ac^A +
\frac{\delta^r \Psi}{\delta\phi^A}(c^A + {\cal{R}}^A[\phi]) ~.
\eeq

The final gauge-fixed partition function is thus again of a simple form
\beq
{\cal{Z}} = \int [d\phi][d\varphi][d\phi^*][dB]\delta\left(\phi^*
- \frac{\delta^r\Psi}{\delta\phi^A}\right)
\exp\left[\frac{i}{\hbar} S_{g.f.}\right] ~.
\eeq
Comparing with eq. (79), the only difference is that after integrating
over the ghost fields $c^A$, the antighosts $\phi^*_A$ no longer
appear in the action. Instead, there is now an extra term multiplying
$\delta\Psi/\delta\phi^a$, and the final answer is of course the same.

Note that the presciption is still to gauge fix by replacing
$\phi^*_A = 0$ by $\phi^*_A = \delta^r\Psi/\delta\phi^A$, but that now
this replacement is entirely trivial since the extended action is
$\phi^*_A$-independent. Instead, the action contains the additional
term
\beq
\delta\Psi[\phi] = \frac{\delta^r\Psi}{\delta\phi^A}{\cal{R}}^A ~,
\eeq
which is just the usual gauge fixing term one adds to the
classical action in standard Lagrangian BRST quantization. This
choice of the extended action corresponds to the boundary condition
for the Master Equation that gives us just the classical action as
the solution after having integrated out the shift ghosts $c^A$.
Of course, this is also a valid boundary condition, but the
disadvantage is that one is back in the standard Lagrangian BRST
formulation and one must then perform further steps in order to
gauge fix the action (by adding an appropriate term of the form
$\delta\Psi[\phi]$). This is, however, all automatically achieved
in one step by the collective field technique.

At the level of the Master Equation, the choice (88) corresponds to
\beq
\frac{1}{2}(S_{ext},S_{ext}) ~=~ -\frac{\delta^r S_{ext}}
{\delta \phi^A}(c^A + {\cal{R}}^A) ~.
\eeq
Since all steps are otherwise identical, the solution to eq. (93)
must be a solution to eq. (85) as well. This shows that whenever we
have a solution $S_{ext}[c^A, \ldots]$, also $S_{ext}[c^A +
{\cal{R}}^A, \ldots]$ is a valid solution. It is trivial to see
that in fact an arbitrary coefficient $\alpha$ can be introduced as
well:
\beq
S_{ext}[c^A,\ldots] \to S_{ext}[c^A + \alpha{\cal{R}}^A,\ldots] ~.
\eeq
This redundancy is nothing but the expression that the solution for the
extended action is also automatically invariant under the internal
BRST symmetry:
\beq
\frac{\delta S_{ext}}{\delta\phi^A}{\cal{R}}^A = 0~.
\eeq

The functional measures should be specified as well. It has been our
working assumption that all measures of the fundamental fields are
flat, $i.e.$ invariant under arbitrary local shifts. If one is forced
to quantize theories with more complicated measures, the symmetries
of those measures determine the transformations from which
Schwinger-Dyson equations follow. The corresponding Master Equation
will then also look different, but the principle of imposing the
Schwinger-Dyson equations at the level of Ward Identities can still
be enforced.

The measure of the fundamental fields is therefore by construction
invariant under the Schwinger-Dyson BRST transformation, and in the
particular case of a flat measure they are invariant under the shifts
(83) used in deriving the Master Equation (85). Invariances of the
measure for the antifields $\phi^*_A$ under their transformation
(83) is required as well. This is because in order for the Ward
Identities to be satisfied, at least the combination of measure and
action must be invariant. Since the action $S_{ext}$ of (82) itself
is invariant,
this is required for the measure too. If one chooses a different
measure, the correct Schwinger-Dyson equations will not be recovered
and the quantization scheme is therefore inconsistent. In the
most common case
of Grassmann-valued antifields $\phi^*_A$, the usual rules of
Berezin integration correspond to functional measures that are
invariant under local shifts of the $\phi^*_A$ fields. It turns out
that such measures will also be invariant under the transformation
(83) in the case of (irreducible) closed gauge algebras. The relevant
Jacobian is
\beq
J = 1 - \frac{\delta^r}{\delta\phi^*_A}\left(\frac{\delta^l S_{ext}}
{\delta\phi^A(x)}\mu\right)~,
\eeq
which actually equals unity due to the trace properties of these
gauge algebras. (The parameter $\mu$ is an anticommuting $c$-number
needed to define true variations from BRST transformations; see the
appendix for the conventions).
This is indeed only a consistency check, because it
follows from our derivation in terms of collective fields that the
measure {\em must} be invariant under this transformation. So
invariance is a priori guaranteed. We shall discuss how to go beyond
this case below.

After having
determined the solution $S_{ext}$ of eq. (85) which satisfies the
proper boundary conditions, both the measure and the action is
therefore in this case by
construction invariant under the BRST symmetry (83). Since it can
be viewed as the result of having integrated out certain
auxiliary fields, it is no surprise that it is not nilpotent
off-shell in general. But the crucial property is that it is
off-shell nilpotent when acting on functions of the fields
$\phi^A$ only.

\subsection{Quantum Master Equations}

We have seen from the collective field method that for closed
irreducible gauge algebras we get an extended action that can
be split into a part independent of the new ghosts $c^A$, and
a simple quadratic term of the form $\phi^*_Ac^A$. Let us, for
reasons that will become evident shortly, denote the part which
is independent of $c^A$ by $S^{(BV)}, i.e.$:
\beq
S_{ext}[\phi,\phi^*,c] = S^{(BV)}[\phi,\phi^*] - \phi^*_Ac^A~.
\eeq

This action is invariant under the transformations
\begin{eqnarray}
\delta \phi^A &=& c^A \cr
\delta c^A &=& 0 \cr
\delta \phi^*_A &=& - \frac{\delta^l S_{ext}}
{\delta \phi^A} ~.
\end{eqnarray}
Moreover, the functional measure is also formally guaranteed to be
invariant in this case. It follows that in this case the Ward
Identities of the kind
$0 = \langle \delta[\phi^*_A F[\phi] \rangle$ are the most general
Schwinger-Dyson equations for the quantum theory defined by the
classical action $S[\phi]$.

But demanding that {\em both} the action $S_{ext}$ {\em and} the
functional measure be invariant under the BRST Schwinger-Dyson
symmetry above is not the most general condition. To derive the
correct Ward Identities we only need that just the {\em
combination} of action and measure is invariant. In this subsection
we want to discuss the more general case in which the set of
transformations (98) still generate a symmetry of the combination
of measure and action, but not of each individually. If we
insist on a solution of the form (97), then the other property
that is required, $\langle c^A\phi^*_B\rangle = -i\hbar\delta^A_B$,
follows automatically.

It thus remains to be found under what conditions the combination
of the action and the measure remain invariant under the BRST
Schwinger-Dyson symmetry. With an action $S_{ext}$ of the form
(97), we get
\begin{eqnarray}
\delta S_{ext} &=& \frac{\delta^r S^{(BV)}}{\delta\phi^A}c^A +
\frac{\delta^r S^{(BV)}}{\delta\phi^*_A}\left(-\frac{\delta^l
S_{ext}}{\delta\phi^A}\right) - \frac{\delta^r(\phi^*_Ac^A)}
{\delta\phi^*_B}\left(-\frac{\delta^l S_{ext}}{\delta\phi^A}
\right) \cr
&=& \frac{\delta^r S^{(BV)}}{\delta\phi^A}c^A +
\frac{\delta^r S^{(BV)}}{\delta\phi^*_A}\left(-\frac{\delta^l
S^{(BV)}}{\delta\phi^A}\right) - (-1)^{\epsilon_A+1}c^A
\left(-\frac{\delta^l S^{(BV)}}{\delta\phi^A}\right) \cr
&=& -\frac{\delta^r S^{(BV)}}{\delta\phi^*_A}\frac{\delta^l
S^{(BV)}}{\delta\phi^A} ~=~ \frac{1}{2}(S^{(BV)},S^{(BV)})~.
\end{eqnarray}

We will still assume that we are integrating over a flat euclidean
measure for the fundamental field $\phi^A$. This measure is formally
invariant under the transformation (98). However, for a corresponding
flat euclidean measure for $\phi^*_A$, the Jacobian of the
transformation (98) will in general be different from unity. As we
already discussed above, the Jacobian equals
\beq
J = 1 - \frac{\delta^r}{\delta\phi^*_A}\left(\frac{\delta^l S_{ext}}
{\delta\phi^A(x)}\mu\right)~.
\eeq

Thus to demand that the combination of measure and action remains
invariant, we must in general require that
\beq
\frac{1}{2}(S_{ext},S_{ext}) =
-\frac{\delta^r S_{ext}}{\delta\phi^A}c^A
+ i\hbar \Delta S_{ext} ~,
\eeq
which, assuming the form (97) -- since we know that this is sufficient
to guarantee the correct Schwinger-Dyson equations -- reduces to
the quantum Master Equation of Batalin and Vilkovisky:
\beq
\frac{1}{2}(S^{(BV)},S^{(BV)}) =  i\hbar \Delta S^{(BV)}~.
\eeq

Let us emphasize that this equation follows even {\em before} possible
gauge fixings. It is required in order that the general
Schwinger-Dyson equations for the fundamental fields are satisfied,
and is not postulated on only the requirement that the final functional
integral be independent of the gauge-fixing function. However, gauge
independence of the functional integral upon the addition of a term
of the form $\delta\Psi[\phi]$ now follows straightforwardly, since
for a functional $\Psi$ that depends only on the fields $\phi$, we
have $\delta^2\Psi[\phi] = 0$.

It may then be worthwhile to leave the actual derivation of the quantum
Master Equation
\beq
\frac{1}{2}(S_{ext},S_{ext}) = -
\frac{\delta^r S_{ext}}{\delta\phi^A}c^A~ + i\hbar\Delta S_{ext},
\eeq
behind. Instead one can, in the spirit of Batalin and Vilkovisky
\cite{Batalin}, take it as the starting point of a more algebraic
approach to quantization. At least two questions should, however,
first be answered. Which are the boundary conditions one should impose
on this equation, and is eq. (97) the most general acceptable
solution? Why does this Master Equation guarantee the correct
quantization prescription?

Consider the last question first. We have seen that the Master
Equation (103) follows from requiring a BRST symmetry of the form
(98). If the solution to the Master Equation contains precisely one
term of the form $\phi^*_Ac^A$ (as in eq. (97), this BRST symmetry
guarantees that the
Schwinger-Dyson equations for all the fundamental fields are
satisfied at the formal level. (It can only be at the formal level,
because without complete gauge fixing, the path integral formalism
is still not totally well-defined.) Having the full set of
Schwinger-Dyson equations satisfied in a well-defined manner
can be viewed as the only independent means of defining what we mean
by a correctly quantized theory.

To ensure that the Schwinger-Dyson equations are satisfied not
only at the formal level, one must therefore specify additional
boundary conditions on
the solution for $S_{ext}$. These boundary conditions must be
imposed in such a way that the path integral becomes well-defined,
without changing gauge-invariant Green functions.
Once the path integral is well-defined, the BRST algebra ensures
that all Schwinger-Dyson equations are satisfied, and the full theory
is then by definition correctly quantized.

When the solution of the equation (103) is chosen to be of the form
(97),
correct boundary conditions can be copied directly from the formalism
of Batalin and Vilkovisky. This follows from the equivalence between
eqs. (101) and (102) in that case. The whole problem has then in effect
reduced to the original Batalin-Vilkovisky formulation.

Imposing the condition that $S_{ext}$ contains only the term
$\phi^*_Ac^A$ and no higher orders in $c^A$ is perhaps not the
only possibility. One way of stating it is that the measures for
$\phi^*_A$ and $c^A$ are not fixed by any overall principle.
A general requirement is that at least
\beq
\langle c^A\phi^*_B \rangle = -i\hbar\delta^A_B ~.
\eeq
This is needed to ensure that the Schwinger-Dyson
equations of the fundamental fields are satisfied. However, the
Schwinger-Dyson equations for the original classical fields
will in general be those associated with an action $S_{ext}$, and
{\em not} those of the classical action $S$. An additional
critereon is then required to select a solution which yields the
correct Schwinger-Dyson equations for the classical fields.
This makes the choice of correct boundary conditions far more
complicated, and we have not investigated this question in
detail. By choosing a solution
containing only the term $\phi^*_Ac^A$ in the action, one is
guaranteed that $\phi^*_A$ is set to zero before gauge fixing. Then
all additional terms in $S_{ext}$ are in fact effectively zero,
and the correct Schwinger-Dyson equations for the classical
fields arise automatically as a consequence of the first
Batalin-Vilkovisky boundary condition. This shows the extent to
which the $\delta$-function ``gauge" for $\phi^*_A$ is the most
general valid condition. (Adding a term of the
form $\delta\Psi$ to the action still yields correct Schwinger-Dyson
equations for gauge invariant objects, since such a term does not
alter gauge invariant Green functions.) In cases where the
derivation can be based straightforwardly on the collective field
technique one can certainly entertain the idea of more complicated
gauge choices for the collective fields -- gauge choices which
could give rise to different valid prescriptions for the
$\phi^*_A$-integrations.

\subsection{Quantum BRST}

In section 2 we noted that the usual BRST Schwinger-Dyson
symmetry acquires a ``quantum correction" if one insists on using
the formalism where the new ghost fields $c^A$ have been integrated
out of the path integral. As we saw already in the case of no gauge
symmetries, this deforms the BRST operator:
\beq
\delta \to \sigma = \delta - i\hbar\Delta ~.
\eeq
The notation is not entirely precise, because the operator $\delta$
on the right hand side of this equation of course equals the operator
$\delta$ on the left hand side only modulo those changes incurred
by integrating out the ghosts $c^A$. But we keep it like this in
order not to clog up the paper with yet more notation. After having
integrated out the ghosts $c^A$, the BRST operator $\delta$ will
become identical to the variation within the antibracket.

Since this quantum deformation involves the same operator $\hbar\Delta$
that in certain specific cases may modify the classical Master Equation,
one might be led to believe that these two issues are related, $i.e.$,
that the ``quantum BRST" operator should only be applied when there are,
(or as a consequence of having) quantum corrections in the full
gauge-fixed action. This is actually not the case, and we
therefore find it useful
to return briefly to the meaning of the quantum BRST operator, here
denoted by $\sigma$.

Let us again choose the simplest solution to the Master Equation of
the form (97). We emphasize that it is immaterial whether this
extended action $S_{ext}$ satisfies the classical or quantum Master
Equations. Since we are interested in seeing the effect of integrating
out the ghosts $c^A$, consider, as in section 2, the expectation
value of the BRST variation of an arbitrary functional
$G = G[\phi^A,\phi^*_A]$:
\begin{eqnarray}
\langle \delta G[\phi,\phi^*] \rangle &=& {\cal{Z}}^{-1}\int
[d\phi][d\phi^*][dc] \delta G[\phi,\phi^*]\exp\left[\frac{i}{\hbar}
\left(S^{(BV)} - \phi^*_Ac^A\right)\right] \cr
&=& {\cal{Z}}^{-1}\int
[d\phi][d\phi^*][dc]\left\{\frac{\delta^rG}{\delta\phi^A}c^A
+ \frac{\delta^rG}{\delta\phi^*_A}\left(-\frac{\delta^l S_{ext}}
{\delta\phi^A}\right)\right\}
\exp\left[\frac{i}{\hbar}\left(S^{(BV)} - \phi^*_Ac^A\right)\right]
\cr
&=& {\cal{Z}}^{-1}\int
[d\phi][d\phi^*]\left\{\frac{\delta^rG}{\delta\phi^A}(i\hbar)
\frac{\delta^l}{\delta\phi^*_A}\delta(\phi^*)
- \frac{\delta^rG}{\delta\phi^*_A}\frac{\delta^l S^{(BV)}}
{\delta\phi^A}\delta(\phi^*)\right\}
\exp\left[\frac{i}{\hbar}S^{(BV)}\right] \cr
&=& {\cal{Z}}^{-1}\int
[d\phi][d\phi^*]\delta(\phi^*)\left\{\frac{\delta^rG}{\delta\phi^A}
\frac{\delta^l S^{(BV)}}{\delta\phi*_A} + (i\hbar)(-1)^{\epsilon_A}
\frac{\delta^r\delta^rG}{\delta\phi*_A\delta\phi^A}
- \frac{\delta^rG}{\delta\phi^*_A}\frac{\delta^l S^{(BV)}}
{\delta\phi^A}\right\} \cr
&& \times \exp\left[\frac{i}{\hbar}S^{(BV)}\right] \cr
&=& \langle (G,S^{(BV)}) - i\hbar\Delta G \rangle~.
\end{eqnarray}
The derivation given here corresponds to the path integral before
gauge fixing, but it goes through in entirely the same manner in
the gauge-fixed case. (The only difference is that the relevant
$\delta$-function reads $\delta(\phi^* - \delta^r\Psi/\delta
\phi)$ instead of $\delta(\phi^*)$; this does not affect the
manipulations above).

The emergence of the ``quantum correction" in the BRST operator
is thus completely independent of the particular solution
$S^{(BV)}[\phi,\phi^*]$; it must always be included when one uses
the formalism in which the ghosts $c^A$ have been integrated out.
The quantum BRST operator $\sigma$ is unusual, because it
appears only after functional manipulations inside the path
integral.

Since by construction the partition function is invariant under
$\delta$ (when keeping the ghosts $c^A$) and $\sigma$ (after
having integrated out these ghosts), it follows that all expectation
values involving these operators vanish:
\beq
\langle \delta G[\phi,\phi^*] \rangle = 0
\eeq
when keeping $c^A$, and
\beq
\langle \sigma G[\phi,\phi^*] \rangle = 0
\eeq
when the $c^A$ have been integrated out.

This of course holds for the action as well:
\beq
\langle \delta S_{ext} \rangle = 0~;~~~~\langle \sigma S^{(BV)}
\rangle =  0~.
\eeq

The first of these equations is trivially satisfied when $S_{ext}$
satisfies the classical Master Equation, because then the variation
$\delta S_{ext}$ itself vanishes. This equation is then only
non-trivially satisfied when $\Delta S_{ext} \neq 0$.

Since the two operations $\delta$ and $\sigma$ are equivalent in
the precise sense given above, the
same considerations should apply to the second equation.
Indeed it does: When $S^{(BV)}$ satisfies the classical Master
Equation, $\sigma S^{(BV)} = 0$ at the operator level, while
that equation is satisfied only in terms of expectation values when
$\Delta S^{(BV)} \neq 0$.

Note that when $\Delta S_{ext} \neq 0$ (or $\Delta S^{(BV)} \neq 0$),
the quantum action is neither invariant under $\delta$ nor $\sigma$.
The action precisely has to remain non-invariant in order to cancel
the non-trivial contribution from the measure in that case. This is
the origin of the factor 1/2 difference between the quantum Master
Equation
\beq
\frac{1}{2}(S^{(BV)},S^{(BV)}) - i\hbar\Delta S^{(BV)} = 0
\eeq
and the operator $\sigma$ (when acting on $S^{(BV)}$):
\beq
\langle (S^{(BV)},S^{(BV)}) - i\hbar\Delta S^{(BV)} \rangle = 0 ~.
\eeq
The combination of these two equations yields the new identities
\beq
\langle \Delta S^{(BV)} \rangle = 0~,~~~ \langle (S^{(BV)},
S^{(BV)}) \rangle = 0 ~
\eeq
which can also be verified directly using the path integral.

The operator $\delta$ defines a BRST cohomology only on the
subspace of fields $\phi^A$; it is only nilpotent on that
subspace. The operator $\sigma$ is nilpotent in general:
$\sigma^2 = 0$ (a consequence of having performed partial
integrations in deriving it). However, the two operators
share the same physical content.

\subsection{Open Gauge Algebras}

Having the quantum Master Equation available, it is
of interest to lift the restriction to closed irreducible
gauge algebras. Recall that in that case the functional
measure is formally BRST invariant, and the classical Master
Equation suffices. Directly related to this is the fact that
this case can be dealt with straightforwardly through the
use of certain collective fields. We believe that the case
of reducible gauge symmetries can be treated
in a rather similar fashion, but something
really new enters when one considers theories with open gauge
algebras. Here the quantum Master Equation seems to be the
preferable starting point.

An open gauge algebra is characterized by closing only on-shell.
Off-shell a new term, proportional to equations of motion
appear. For this reason, the collective field formalism has to be
modified in order to be valid at the quantum level. This means that
it is not straightforward to introduce collective fields so as to
enforce well-defined Schwinger-Dyson equations in that case.

Let us thus first follow the reasoning above, and view the Master
Equation (103) as the key to the solution of this problem,
independently of its
derivation by means of collective fields. We know that this point of
view is acceptable, because it automatically guarantees correct
Schwinger-Dyson equations.

Some notation should be introduced. The gauge generators defined by
the symmetry
\beq
\delta \phi^A = R^A_{\alpha}\epsilon^{\alpha}
\eeq
$i.e.$ through
\beq
\frac{\delta S}{\delta\phi^A}R^A_{\alpha} = 0
\eeq
for the classical fields $\phi^A$, form (see, $e.g.$ ref. \cite{Bat1})
an algebra
\beq
\frac{\delta R^A_{\alpha}}{\delta\phi^B}R^B_{\beta} - \frac{\delta
R^A_{\beta}}{\delta\phi^B}R^B_{\alpha} = - R^A_{\gamma}
f^{\gamma}_{\alpha\beta} - \frac{\delta S}{\delta\phi^B}E^{AB}_
{\alpha\beta} ~.
\eeq
We have for convenience restricted ourselves to the case of bosonic
fields $\phi^A$. The coefficients $E^{AB}_{\alpha\beta}$ can be
considered new additional generators of the algebra. For closed gauge
algebras all $E^{AB}_{\alpha\beta}$ vanish.

The solution to the Master Equation (103) can, if we again restrict
ourselves to solutions of the form (97), be immediately read off
from the solution to the quantum Master Equation of Batalin
and Vilkovisky \cite{Batalin,Bat1}.
The result for the extended action is
\beq
S_{ext} = S[\phi] - \phi^*_A(c^A - {\cal{R}}^A) +
\frac{1}{4}\phi^*_A\phi^*_BE^{AB}_{\alpha\beta}c^{\alpha}c^{\beta}~,
\eeq
where $c^{\alpha}$ is the usual ghost associated with the symmetry
(113). To this must be added the quantum corrections.
Open gauge algebras of higher degree can be treated similarly
\cite{Batalin}.

By adding an appropriate gauge-fixing function $\Psi$, and integrating
over the ghost-antighost pairs $\phi^*_A, c^A$, we of course
recover the known result. It is again important that $\Psi$ is a
function of just the fields $\phi^A$, since only on that subspace is
the BRST Schwinger-Dyson operator $\delta$ nilpotent. This is also
the condition which ensures that in the process of gauge-fixing
the antifields are changed from zero to $\delta^r\Psi/\delta\phi^A$.

Although one can demand correct Schwinger-Dyson equations
by construction in this
manner, it is interesting that the standard collective field technique
breaks down in this case. If we trace it back, we note that this is
because demanding BRST invariance of the classical action by itself
is too restrictive, and in fact not required. We only need to insist
on correct Schwinger-Dyson equations for the full gauge-fixed theory.
For open gauge algebras the transformations of the fields and their
collective field partners should therefore not necessarily be required
to leave the classical action
invariant. Since at the level before integrating out the collective
fields $\varphi^A$ and their Nakanishi-Lautrup fields $B_A$ one should
have a full nilpotent BRST algebra, one can in fact derive the required
transformation laws. This is highly analogous to the original general
solution of De Wit and van Holten \cite{Wit}. We believe that the
correct procedure for introducing collective fields in this case
could go along such lines, but we have not investigated this question
in detail. It is also possible that it may require the introduction
of additional fields, perhaps resulting from the gauging of ``trivial"
gauge symmetries related to the open algebra structure coefficients,
a point of view that has been advocated by Hull \cite{Hull}. In any
case, we find it quite likely that a consistent interpretation in terms
of suitable collective fields exists in this more general case as well.

Finally a few words about the quantum corrections to the gauge-fixed
action. These may seem unusual from the point of view of more
conventional BRST gauge fixing, where it ordinarily suffices to add
to the classical action a term of the form $\delta\Psi$, with
$\delta$ being the variation with respect to a nilpotent BRST
operator. This, however, assumes that the functional measure remains
invariant under the transformation. If it does not, then the BRST
procedure of, $e.g.$, ref. \cite{Wit} must be supplemented. One can do
this by trying to find a term whose BRST variation precisely
equals the contribution from the measure. If this is not possible
(because adding the extra term may modify the Jacobian),
one must resort to a higher-order expansion in $\hbar$, at each order
trying to correct for the change in the measure. This is the ordinary
BRST quantization analogue of solving the general quantum Master
Equation in the Batalin-Vilkovisky framework.

\section{Conclusion}

We have shown that a new set of ghost fields, $c^A$ in the notation
of this paper, naturally belong to the Batalin-Vilkovisky Lagrangian
quantization scheme. What in the language
of Batalin and Vilkovisky are known as ``antifields", are the usual
BRST partners (antighosts) of these new ghost fields.

The ghosts $c^A$ appear when one insists that Schwinger-Dyson equations
should be satisfied at the level of BRST Ward Identities. After
integrating out these ghosts $c^A$ one recovers the full scheme of
Batalin-Vilkovisky: The antibracket, the quantum Master Equation,
the quantum BRST generator etc. It is truly remarkable that
almost the whole formalism could be developed \cite{Batalin} without
this additional background, and without a corresponding derivation
from more known quantization principles. The only other clue to
this formalism that we know of is the Zinn-Justin equation for
Yang-Mills theory \cite{Zinn}.

Interestingly, the fundamental BRST symmetry for any quantum field
theory of flat measures for the fields is not the usual BRST symmetry
associated with (possible) internal gauge invariances, but rather
\begin{eqnarray}
\delta \phi^A &=& c^A \cr
\delta c^A &=& 0 \cr
\delta \phi^*_A &=& - \frac{\delta^l S_{ext}}{\delta\phi^A} ~.
\end{eqnarray}
This BRST symmetry, which we have called the BRST Schwinger-Dyson
symmetry contains {\em no} explicit reference to the internal gauge
(or internal BRST) invariances among the fields $\phi^A$. Instead,
all knowledge of the gauge transformations has been transferred into
certain conditions the extended action $S_{ext}$ must fulfil. This
corresponds to choosing appropriate boundary conditions for the
solution to the Master Equation.

Only when we integrate out the ghost-antighost pair $c^A,\phi^*_A$
do we recover the usual internal BRST transformation for the fields
themselves. In the simplest cases, these internal BRST symmetries
are nothing but the
equations of motion for the new ghost fields. Intuitively we can
understand this late appearance of the
standard internal BRST algebra from the fact that Schwinger-Dyson
equations represent much more general statements about the quantum
theory, above such details as the particular symmetries that
leave the theory invariant.

A new Master Equation follows if one keeps the new ghosts. In those
cases where we can derive the quantization scheme straightforwardly
from collective fields, it is of the simple form
\beq
\frac{1}{2}(S_{ext},S_{ext}) =
-\frac{\delta^r S_{ext}}{\delta\phi^A}c^A
\eeq
with no quantum corrections. These cases include all theories without
internal gauge symmetries, and theories invariant under transformations
satisfying a closed irreducible algebra. The functional measures are in
those cases formally invariant.

When one cannot immediately derive the correct
prescription through the use of collective fields, one can instead
rely on the more general principle of satisfying Schwinger-Dyson
equations in the quantum theory. The only input is then the
Schwinger-Dyson BRST symmetry above, and one must now demand that at
least the combination of measure and action remains invariant under
this symmetry (since this suffices to derive the Schwinger-Dyson
equations as Ward Identities of this symmetry). For flat
$\phi^A$-measures, the relevant Jacobian differs from unity if and
only if
$\Delta S_{ext} \neq 0$. In that case the correct quantum action
$S_{ext}$ will not be invariant under the BRST symmetry (117), but
will instead transform in precisely such a manner as to cancel the
Jacobian from the measure. The Master Equation (118) must then be
replaced by the quantum Master Equation
\beq
\frac{1}{2}(S_{ext},S_{ext}) =
-\frac{\delta^r S_{ext}}{\delta\phi^A}c^A
+ \hbar \Delta S_{ext} ~.
\eeq
This equation follows directly from demanding that the Schwinger-Dyson
equations are Ward Identities of the symmetry (117). The quantum
Master Equation may also be required in situations where consistent
regulators that preserve the relevant BRST symmetry cannot be found.
This may be the case for anomalous theories \cite{Troost}, or in
larger generality. An example of how this may lead to ``anomalous
gauge fixing" has been given in ref. \cite{Bos} in more
conventional BRST language, but it presumably
has a corresponding interpretation in terms of the
Batalin-Vilkovisky quantum Master Equation.

The ``quantum deformation" of the BRST generator from $\delta$ to
$\sigma = \delta - i\hbar \Delta$ is unrelated to the need in
certain cases for having a quantum correction to the Master
Equation, as in eq. (119) above. The quantum Master Equation is only
required in those cases where the functional measure does not
respect the BRST Schwinger-Dyson algebra. In contrast, the BRST
symmetry operator  is {\em always} $\sigma$ rather than $\delta$
if one insists on the formulation in which the ghosts $c^A$ have
been integrated out, but the antighosts $\phi^*_A$ are kept. If
one keeps the new ghosts $c^A$, there is no quantum deformation,
and the usual BRST operator $\delta$ generates a genuine symmetry
of the path integral. Stated differently, the ``quantum corrections"
to the BRST operator are already automatically included in the
formulation in which the ghosts $c^A$ are kept. In particular,
the Ward Identities obtained
using this operator are the correct Schwinger-Dyson equations.

The simple interpretation of the antifields as antighosts
introduced through the gauge fixing of a certain
Schwinger-Dyson gauge symmetry is lost if one relies solely on the
Master Equation to solve the problem. However, we find it quite
likely that some modification of the usual collective field
technique can be used in general to describe the $\phi^*$-fields as the
antighosts required to fix the relevant gauge symmetry. To get to that
stage will perhaps require the introduction of more field
variables in cases of, $e.g.$, open gauge algebras.

Our discussion has at no point touched the issue of field theoretic
unitarity. It has been shown that when a certain locality condition
\cite{Henneaux} is imposed on the solution to the Master Equation,
unitarity can be guaranteed \cite{Barnish}.

{}From the present point of view, the antifields $\phi^*_A$ are not
artificial devices with which to set up the correct quantization
procedure. They are genuine fields of the path integral, on equal
footing with $c^A$ and all the usual fields $\phi^A$. For the most
straightforward gauge fixings we have considered here, these
ghost-antighost pairs $c^A, \phi^*_A$ do not propagate (and for
this reason are often easy to integrate out of the path integral).
Depending on the circumstances, this can either be seen as
corresponding to very well-behaved
functional integrals, or as having to deal with rather singular
propagators of the $\delta$-function kind. Regulators can be
introduced which preserve the Schwinger-Dyson equations, but which
give rise to propagating fields $c^A$ and $\phi^*_A$. A simple
one-dimensional example which readily generalizes to higher
dimensions has been given in ref. \cite{us2}. As
expected, the corresponding BRST Schwinger-Dyson algebra is,
however, modified (``regularized"). The change in the BRST
transformations can be straightforwardly derived by the collective
field technique.

Inclusion of the new ghosts $c^A$ may be aesthetically appealing, but
will they be helpful in developing the correct quantization
prescription in cases that are not yet fully understood? String theory
clearly comes to mind here. What is the r\^{o}le of these new ghosts
in string field theory? Do they simplify the formalism? Recently
Verlinde \cite{Verlinde} has shown how to derive an equation very
reminiscent of the Batalin-Vilkovisky quantum Master Equation for
low-dimensional string theory. The derivation was noted to be almost
identical to the derivation of Virasoro and $W$-constraints on the
string partition functions. This is presumably no coincidence, since
in both cases they express Schwinger-Dyson equations. Is
it advantageous to include the analogues of the ghosts $c^A$ in
this formulation?

The shift symmetries introduced through collective fields to ensure
Schwinger-Dyson equations at the level of the BRST algebra are only
very special cases of more general field-enlarging transformations
that can be performed within the Feynman path integral. Since the
particular choice of variables should have no influence on physical
quantities, $i.e.$ in this context $S$-matrix elements, one should
be able to formulate the quantization prescription in a more
coordinate-independent manner. This is the content of the field
redefinition theorem. Enforcing Schwinger-Dyson equations in
different field variables can be accomplished by suitable field
transformations, for which the simple shift (and subsequent gauging
to zero of the collective fields) that we have considered in this
paper corresponds to the identity transformation. Switching between
one set of fields to another may, independently of whether one uses
a Lagrangian or Hamiltonian formulation, involve the addition of
terms of up to order $\hbar^2$ to the naively transformed action,
but also these additional terms have a corresponding interpretation
in the BRST Schwinger-Dyson algebra \cite{us2}.

When the functional measures for the fundamental fields are
non-trivial, the Master Equations will differ in form. The advantage
of the present derivation is that we now know how to write down
the corresponding Master Equations for theories of arbitrary
path integral measures.

\vspace{0.5cm}
\noindent
{\sc Acknowledgement:}
The work of J.A. has been partially supported by
Fundaci\'on Andes no. C-11666/4, and an EC CERN Fellowship.

\vspace{2cm}

\appendix{{\Large {\bf Appendix}}}

\vspace{0.3cm}

In this appendix we give some additional conventions, and
list some useful identities.

Repeatedly in this formalism one needs the notion of left and
right derivatives. For bosons only, this notion is not useful.
But as soon as the Grassmann parity of the fields is kept arbitrary,
it pays to use both left and right derivatives in appropriate
places, in order to simplify expressions. The Leibniz rules
for derivations of the left and right kind read
\beq
\frac{\delta^l(F\cdot G)}{\delta A} = \frac{\delta^lF}{\delta A} G
+ (-1)^{\epsilon_F\cdot\epsilon_A} F \frac{\delta^lG}{\delta A}
\eeq
and
\beq
\frac{\delta^r(F\cdot G)}{\delta A} = F \frac{\delta^rG}{\delta A}
+ (-1)^{\epsilon_G\cdot\epsilon_A} \frac{\delta^rF}{\delta A}G ~,
\eeq
where $A$ denotes a field (or antifield) of arbitrary Grassmann
parity $\epsilon_A$. Similarly, $\epsilon_F$ and $\epsilon_G$ are
the Grassmann parities of the functionals $F$ and $G$.

Actual variations, let us denote them by $\bar{\delta}$ in contrast
to the BRST transformations $\delta$ of the paper, are defined as
follows:
\beq
F[A+\bar{\delta}A] - F[A] \equiv \bar{\delta}F \equiv
\bar{\delta}A\frac{\delta^lF}{\delta A} \equiv \frac{\delta^rF}
{\delta A}\bar{\delta}A ~.
\eeq

The commutation rule of two arbitrary fields is
\beq
A\cdot B = (-1)^{\epsilon_A\cdot\epsilon_B}B\cdot A ~,
\eeq
and for actual variations one has the simple rule that
\beq
\bar{\delta}(F\cdot G) = (\bar{\delta}F)G + F(\bar{\delta}G)
\eeq
independent of the Grassmann parities $\epsilon_F$ and $\epsilon_G$.
The rules (122) and (123) in conjunction lead to the useful identity
\beq
\frac{\delta^l F}{\delta A} = (-1)^{\epsilon_A(\epsilon_F+1)}
\frac{\delta^r F}{\delta A} ~.
\eeq

The BRST variations we have worked with in this paper correspond to
right derivation rules. This is of course not imposed upon us, but it
is convenient if we wish to compare our expressions with those of
Batalin and Vilkovisky. It follows from requiring the actual
variations to be related to the BRST transformations by multiplication
of an anticommuting parameter $\mu$ {\em from the right}. This then
provides us with very helpful operational rules for the BRST
transformations $\delta$. In particular,
\beq
\bar{\delta}F \equiv (\delta F)\mu = \frac{\delta^r F}{\delta A}
\bar{\delta} A ~.
\eeq
Now, since
\beq
\delta F \equiv \frac{\delta^r \bar{\delta} F}{\delta\mu} ~,
\eeq
it follows that
\beq
\delta F = \frac{\delta^r F}{\delta A} \delta A~.
\eeq
{}From this it also follows directly that the BRST transformations
act as right derivations:
\beq
\delta(F\cdot G) = F(\delta G) + (-1)^{\epsilon_G}(\delta F)G ~.
\eeq

These are the basic rules that are needed for the manipulations in
the main text.

\end{document}